\documentclass[aps,prmaterials,twocolumn,floats,amsmath,amssymb,superscriptaddress,noeprint,longbibliography]{revtex4-2}

\usepackage{here}
\usepackage{subcaption}
\usepackage{tabularx}
\usepackage{graphicx}
\usepackage{bm}
\usepackage{dcolumn}
\usepackage{tikz}
\usepackage{siunitx}
\usepackage{hyperref}
\usepackage{booktabs}
\usepackage{dcolumn}
\usepackage{color}

\newcommand{\eg}{{e.g.}, }
\newcommand{\ie}{{i.e.}, }
\DeclareSIUnit\atm{atm}

\hypersetup{
  colorlinks,
  citecolor=blue,
  linkcolor=blue,
  urlcolor=blue}

\newcolumntype{d}[1]{D{.}{.}{#1}}
\newcommand{\mc}[1]{\multicolumn{1}{c}{#1}}

\DeclareSIUnit\atm{atm}

\begin{document}

\title{$\beta$-Ga$_2$O$_3$(001) surface reconstructions from first principles and experiment}
\author{Konstantin Lion}
\email[Contact author: ]{lion@ms1p.org}
\thanks{Current address: Molecular Simulations from First Principles e.V., Berlin, Germany}
\affiliation{Institut f\"ur Physik and CSMB, Humboldt-Universit\"at zu Berlin, Berlin, Germany}
\affiliation{\textcolor{black}{The Fritz-Haber-Institut der Max-Planck-Gesellschaft, Berlin, Germany}}
\author{Piero Mazzolini}
\thanks{Current address: Department of Mathematical Physical and Computer Sciences, University of Parma, Parma, Italy}
\affiliation{Paul-Drude-Institut für Festkörperelektronik, Leibniz-Institut im Forschungsverbund Berlin e.V.,Berlin, Germany}
\author{Kingsley Egbo}
\affiliation{Paul-Drude-Institut für Festkörperelektronik, Leibniz-Institut im Forschungsverbund Berlin e.V.,Berlin, Germany}
\author{Toni Markurt}
\affiliation{Leibniz-Institut für Kristallzüchtung (IKZ), Berlin, Germany}
\author{Oliver Bierwagen}
\affiliation{Paul-Drude-Institut für Festkörperelektronik, Leibniz-Institut im Forschungsverbund Berlin e.V., Berlin, Germany}
\author{Martin Albrecht}
\affiliation{Leibniz-Institut für Kristallzüchtung (IKZ), Berlin, Germany}
\author{Claudia Draxl}
\affiliation{Institut f\"ur Physik and CSMB, Humboldt-Universit\"at zu Berlin, Berlin, Germany}

\date{\today}

\begin{abstract}
	We present a comprehensive investigation of reconstructions on $\beta$-Ga$_2$O$_3$(001) combining first-principles calculations with experimental observations. Using  {\it ab initio} atomistic thermodynamics and replica-exchange grand-canonical molecular dynamics simulations, we explore the configurational space of possible reconstructions under varying chemical potentials of oxygen and gallium. Our calculations reveal several stable surface reconstructions, most notably a previously unreported 1$\times$2 reconstruction consisting of paired GaO$_4$ tetrahedra that exhibits remarkable stability across a wide range of experimental growth conditions. In this reconstruction, two Ga atoms share one oxygen bond and are separated by a distance of \SI{2.64}{\angstrom} along the [010] direction. High-angle annular dark-field scanning transmission electron microscopy imaging of homoepitaxially grown (001) layers is consistent with the predicted structure. Additional investigations of possible indium substitution at the surface sites, \textcolor{black}{which can occur during indium-mediated metal-exchange catalysis via molecular beam epitaxial growth}, reveal a cooperative effect in In incorporation, with distinct stability regions for In-substituted structures under O-rich conditions. Our findings provide an understanding for controlling surface properties during epitaxial growth of $\beta$-Ga$_2$O$_3$(001).
\end{abstract}

\pacs{}
\maketitle

\section{Introduction}

Ga$_2$O$_3$, an ultrawide-bandgap transparent semiconducting oxide, has emerged as one of the most promising materials for new-generation power electronics~\cite{green2022,higashiwaki2018,higashiwaki2020,pearton2018}. Its tunable electrical and optical properties make it a promising material for gas sensors~\cite{huang2006,arnold2009,cuong2009,liu2008,mazeina2010}, field-effect transistors~\cite{chang2005}, and photodetectors~\cite{feng2006,li2011,kokubun2007,weng2010}. It can be successfully grown with high quality from the melt~\cite{kuramata2016,galazka2017}, which enables the production of cost-effective bulk crystals. While the material exhibits polymorphism and can adopt at least five different known structures ($\alpha$, $\beta$, $\gamma$, $\delta$, and $\epsilon$)~\cite{roy1952}, most efforts have been concentrated on its thermodynamically most stable phase, $\beta$-Ga$_2$O$_3$. 

The availability of high-quality bulk crystals has enabled systematic investigations of the material's surface properties under controlled conditions. For device applications, understanding and controlling the surface properties during epitaxial growth is crucial. While growth on the (010) surface of $\beta$-Ga$_2$O$_3$ exhibits the highest growth rates due to its high surface free energy~\cite{schewski2018}, its instability under metal-rich conditions~\cite{mazzolini2019a} and the difficulty of preparing (010)-oriented substrates~\cite{blevins2019} make other orientations more attractive. Recent studies have shown that homoepitaxial growth on (001)-oriented substrates can achieve comparable growth rates and structural quality to (010) when using In-mediated metal-exchange catalysis (MEXCAT)~\cite{mazzolini2019a}, where an additional In-flux is added during molecular beam epitaxy (MBE) growth. This makes the (001) direction highly promising for future devices. However, a comprehensive understanding of the (001) surface reconstructions and their stability under realistic growth conditions remains elusive. Previous studies have mainly focused on bulk-truncated surface terminations~\cite{bermudez2006,schewski2018,mu2020}, while the exploration of possible reconstructions, especially under the metal-rich conditions typical of MBE, has been limited.

In this work, we systematically investigate reconstructions of the $\beta$-Ga$_2$O$_3$(001) surface using a combination of first-principles calculations and experimental observations. By employing \textit{ab initio} atomistic thermodynamics~\cite{scheffler1988,qian1988,reuter2001,reuter2003,reuter2005a,rogal2007} together with replica-exchange grand-canonical molecular dynamics simulations~\cite{zhou2019,zhou2020,zhou2022}, the configurational space of possible reconstructions is explored as functions of oxygen and gallium chemical potentials at temperatures relevant for epitaxial growth. We demonstrate how different surface reconstructions can form and validate our predictions through high-angle annular dark-field scanning transmission electron microscopy (HAADF-STEM) imaging of homoepitaxially grown (001) layers. Furthermore, we investigate \textcolor{black}{the role of In on the (001) surface during MEXCAT growth}.

\section{Methods}

\subsection{\textit{Ab initio} atomistic thermodynamics}
\label{sec:aiAT_theory}

The surface free energy ($\gamma$) of a slab, which is influenced by temperature ($T$) and the partial pressures ($p_{i}$) of its constituent elements, is determined as~\cite{scheffler1988,qian1988,reuter2001,reuter2003,reuter2005a,rogal2007}
\begin{align}
    \gamma\left(T,\{p_{i}\}\right) = \frac{1}{2A} \left[G_{\text{slab}}(T,\{p_{i}\}) - \sum_i N_i \mu_i\left(T,p_{i}\right)\right] \, , \label{eq:basic_gamma}
    \end{align}
where $G_{\text{slab}}$ is the Gibbs free energy of the slab, $N_i$ is the number of atoms of element $i$ in the slab, and $\mu_i$ the respective chemical potential. The factor 2 in the denominator indicates that we are simulating symmetric slabs with equal surfaces at \textcolor{black}{their} top and bottom.

The thermodynamic stability of a Ga$_2$O$_3$ slab in contact with reactive gallium and oxygen gas phases \textcolor{black}{is} evaluated in a "constrained" equilibrium approach~\cite{reuter2001, reuter2003}, where surfaces are considered to be in full equilibrium with the oxygen and gallium reservoirs, which are mutually independent of each other. The thermodynamical stability of the slab is then evaluated by comparing its surface free energy to that of a reference slab~\cite{scheffler1988,qian1988,reuter2001,reuter2003,reuter2005a,rogal2007}:
\begin{equation} 
\begin{split}
    \Delta \gamma &(T,p_{\text{O}},p_{\text{Ga}}) = \frac{1}{2A} \bigg[  G_{\text{slab}}(T,p)-G_{\text{ref}}(T,p) \\
    & -\Delta N_{\text{O}}\,\mu_{\text{O}}( T,p_{\text{O}} )-\Delta N_{\text{Ga}}\,\mu_{\text{Ga}} ( T,p_{\text{Ga}}) \bigg] \, .
\end{split}
\label{eq:delta_gamma}
\end{equation}
Here, the differences in the number of oxygen and gallium atoms in the two slabs are represented by $\Delta N_{\text{O}}=N_{\text{O}}^{\text{slab}}-N_{\text{O}}^{\text{ref}}$ and $\Delta N_{\text{Ga}}=N_{\text{Ga}}^{\text{slab}}-N_{\text{Ga}}^{\text{ref}}$, and the chemical potentials of gallium and oxygen are $\mu_{\text{Ga}}$ and $\mu_{\text{O}}$, respectively. In this work, we have chosen the most stable bulk-truncated (001) surface, (001)-B, as the reference system. The Gibbs free energies are approximated by density-functional theory~(DFT) total energies, as commonly done in the literature~\cite{scheffler1988,qian1988,reuter2001,reuter2003,reuter2005a,rogal2007}, at the levels of the generalized gradient approximation (GGA) and hybrid DFT, $G \approx E^{\text{DFT}}$. We will, however, explicitly show for the two most stable terminations that including the vibrational free energy $F_{\text{vib}}$ can influence the relative stability between surfaces on the order of several \si{\milli\electronvolt\per\angstrom^2}. Details of the \textit{ab initio} calculations can be found in \textcolor{black}{Appendix}~\ref{sec:comp_details}.

When considering the free energy, we get an additional term that contributes to $\Delta \gamma$:
\begin{equation} 
\begin{split}
    \Delta \gamma_{\text{vib}} & (T) = \frac{1}{2A} \bigg[ F^{\text{vib}}_{\text{slab}}(T)-F^{\text{vib}}_{\text{ref}}(T) \\
   & -\Delta N_{\text{O}}\,F^{\text{vib}}_{\text{O}}(T)-\Delta N_{\text{Ga}}\,F^{\text{vib}}_{\text{Ga}}(T) \bigg] \, .
\end{split}
\label{eq:delta_gamma_vib}
\end{equation}
The chemical potentials are referenced to the standard states of the elements, \ie the metallic Ga bulk, $\mu_{\text{Ga}}(T,p_{\text{Ga}})=\Delta\mu_{\text{Ga}}(T,p_{\text{Ga}}) + E_{\text{bulk}}^{\text{Ga}}$, and molecular O$_2$, $\mu_{\text{O}}(T,p_{\text{O}})=\Delta\mu_{\text{O}}(T,p) + 1/2 E_{\text{O}_2}$, with the energies \textcolor{black}{including} the zero-point contributions. 
According to~\cite{reuter2003,rogal2007}, we convert $\Delta\mu(T,p)$ into pressure and temperature conditions using thermodynamical tables~\cite{stull1971},
\begin{equation}
    \Delta\mu(T,p) = \Delta\mu(T,p^0) + \frac{1}{2} k_\text{B} T \ln\left(\frac{p}{p^0}\right) \, ,
\end{equation}
where $k_\text{B}$ is the Boltzmann constant and $p^0$ ambient pressure (\SI{1}{\atm}). The bounds for the chemical potentials can be deduced from the condition of the thermodynamic stability of bulk Ga$_2$O$_3$. Below the O-poor (Ga-rich) limit, the oxide will decompose into Ga metal and oxygen, while in the O-rich (Ga-poor) limit, oxygen will condense on the surface. Reasonable bounds for the chemical potentials are thus given as~\cite{reuter2001}
\begin{align}
    \frac{1}{3}H_f(T=\SI{0}{\kelvin},p=0)&< \Delta\mu_{\text{O}} < 0 \, , \\
    \frac{1}{2}H_f(T=\SI{0}{\kelvin},p=0)&< \Delta\mu_{\text{Ga}} < 0 \, ,
\end{align}
where $H_f(T=\SI{0}{\kelvin},p=0)$ is the enthalpy of formation for $\beta$-Ga$_2$O$_3$, for which we obtain \SI{-9.64}{\electronvolt} using PBEsol and \SI{-9.98}{\electronvolt} using PBE0 with a mixing factor $\alpha=0.26$, termed PBE0(0.26) (see \textcolor{black}{Appendix}~\ref{sec:comp_details} for an explanation for this choice). This agrees reasonably well with the experimental value at ambient conditions, which is \SI{-11.29}{\electronvolt}~\cite{lideed.}. The limits for the chemical potentials are thus~$-3.21 \, [-3.33]\,\si{\electronvolt}< \Delta\mu_{\text{O}} < \SI{0}{\electronvolt}$ and~$-4.82 \, [-4.99]\,\si{\electronvolt}< \Delta\mu_{\text{Ga}} < \SI{0}{\electronvolt}$ for PBEsol~[PBE0(0.26)]. When depicting phase diagrams and free surface energies, we will include a range slightly outside these limits, namely $-4.0\,\si{\electronvolt}< \Delta\mu_{\text{O}} < \SI{0}{\electronvolt}$ and $-6.0\,\si{\electronvolt}< \Delta\mu_{\text{Ga}} < \SI{0}{\electronvolt}$. \textcolor{black}{This is not only better comparable to the experimental values, but also accounts for the fact that in reality the enthalpy of formation is temperature and pressure dependent and that the true range of chemical potentials is determined by $H_f(T,p)$~\cite{reuter2001}.}

\subsection{Selection of surface structures}
\label{sec:selection}


\begin{figure}[hbt]
\includegraphics[width=0.45\textwidth]{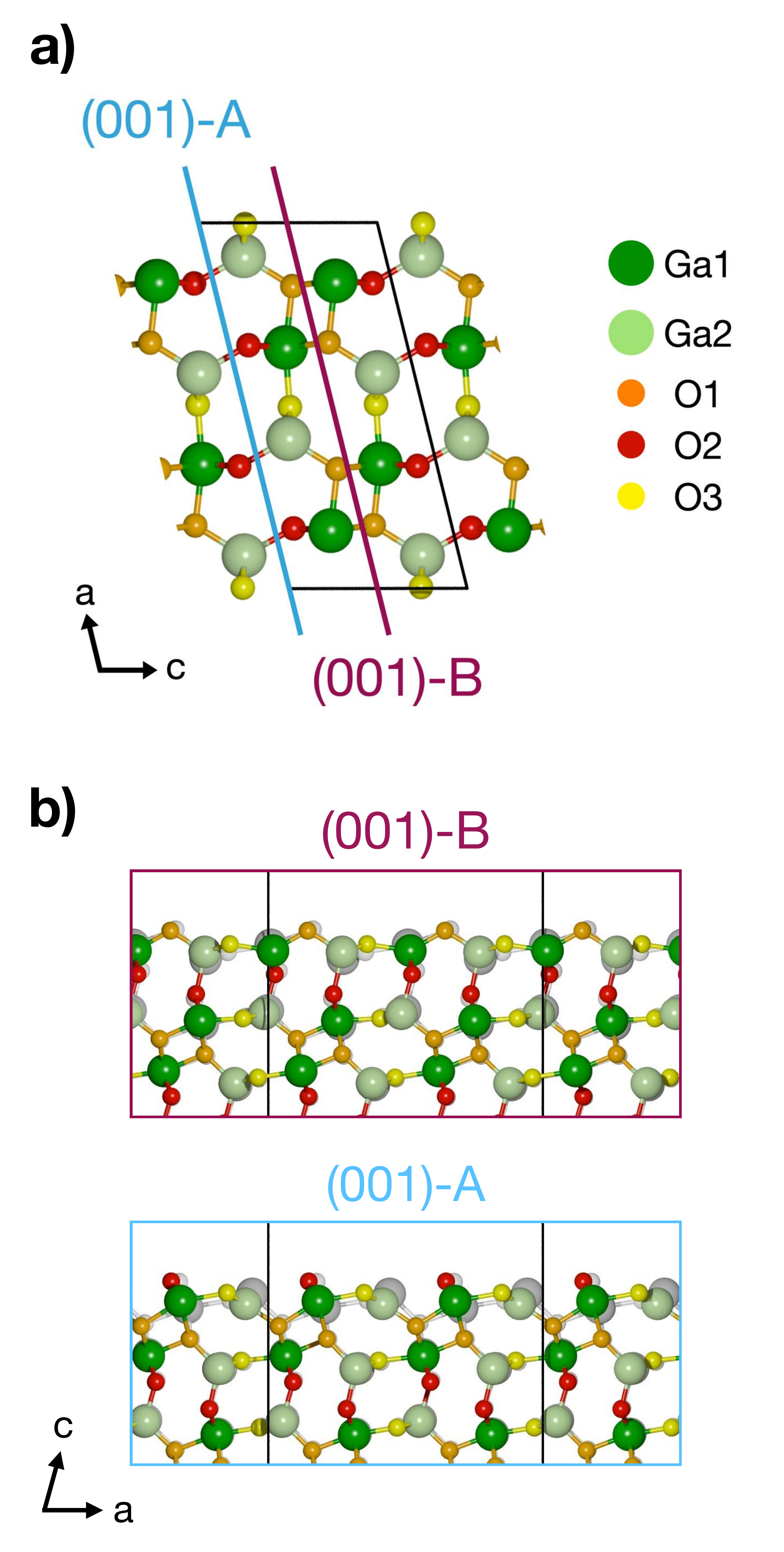}
\caption{(a) Conventional unit cell of $\beta$-Ga$_2$O$_3$ viewed along [010]. The octahedrally (tetrahedrally) coordinated gallium atoms, Ga1 (Ga2), are dark (light) green. The inequivalent oxygen atoms are shown in orange (O1), red (O2), and yellow (O3). The stoichiometric (001)-A and (001)-B terminations are indicated by colored lines. (b) Overlays of the unrelaxed (gray) and relaxed (bright color) stoichiometric (001)-A and (001)-B surface terminations. Shown is a side view of the surface structures projected along [010].}
\label{fig:bulk_structure} \label{fig:001_A_B_truncated_structures}
\end{figure} 

The {\it ab initio} atomistic thermodynamics (aiAT) approach is very powerful, but requires the explicit consideration of all possibly relevant structures. Obviously, it carries the risk that a relevant structure is overlooked. It is therefore necessary to put effort into selecting suitable structures before calculating phase diagrams from aiAT. In this work, we combine several approaches to obtain a suitable set of (001) surface structures as input to the aiAT formalism. All of these surface structures are obtained as adsorbate structures based on the (001)-B or (001)-A termination. This means that while some structures may be interpreted (or are actually formed) as, \eg vacancy structures based on the B (or A termination), they will be treated as adsorbate structures in the following. Pathways between metastable structures are also not explored, since we are mainly interested in the stable equilibrium structures.

\begin{figure}[htb]
\centering
\includegraphics[width=0.5\textwidth]{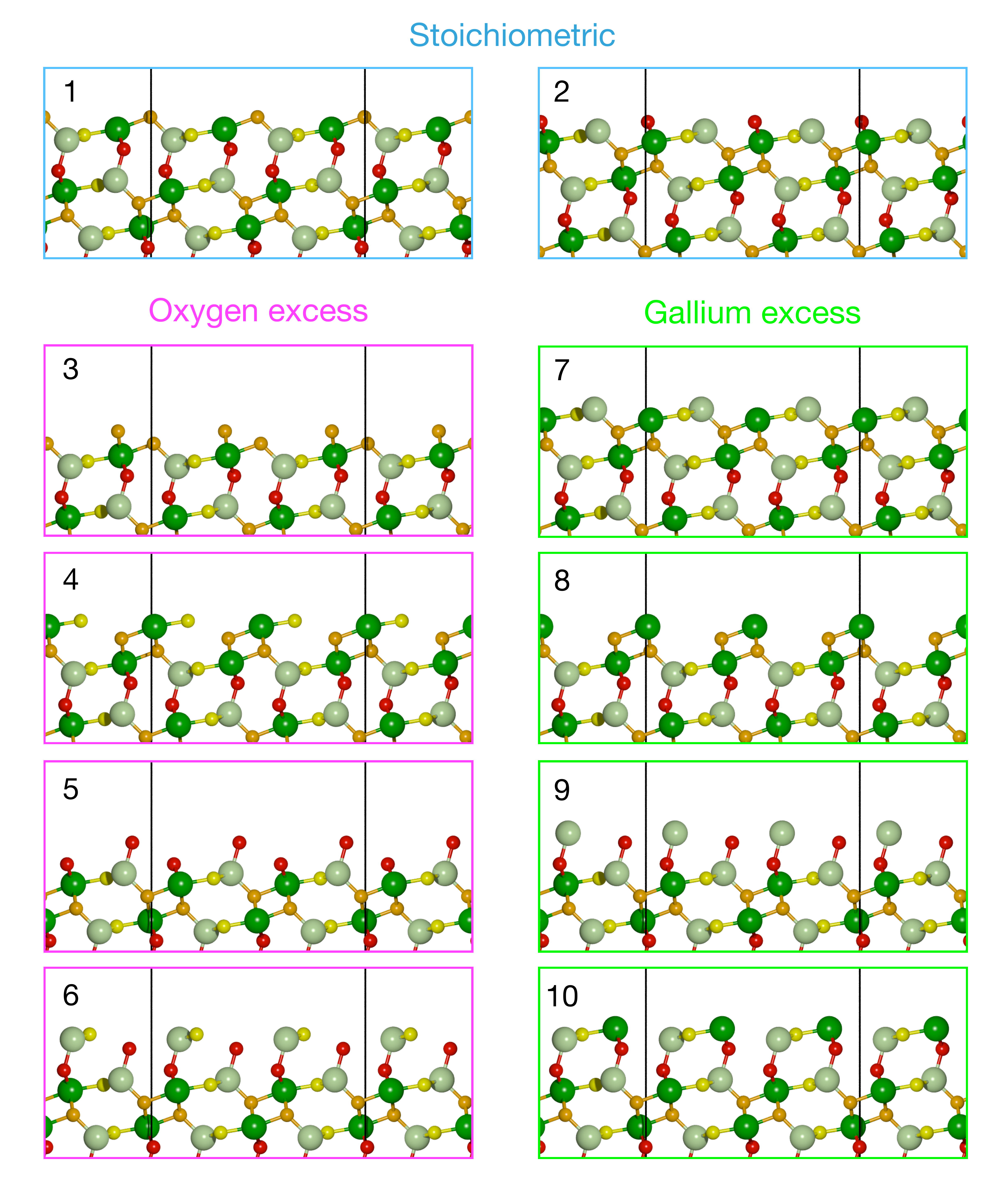}
\caption{Unrelaxed bulk-truncated (001) surface terminations. The stoichiometric terminations are (001)-B~(left) and (001)-A~(right). The different terminations are numbered for easier referencing.} \label{fig:bulk_truncated_structures}
\end{figure}

\begin{figure}[hbt]
\centering
\captionsetup[subfigure]{font=small,labelfont=small}
\includegraphics[width=0.45\textwidth]{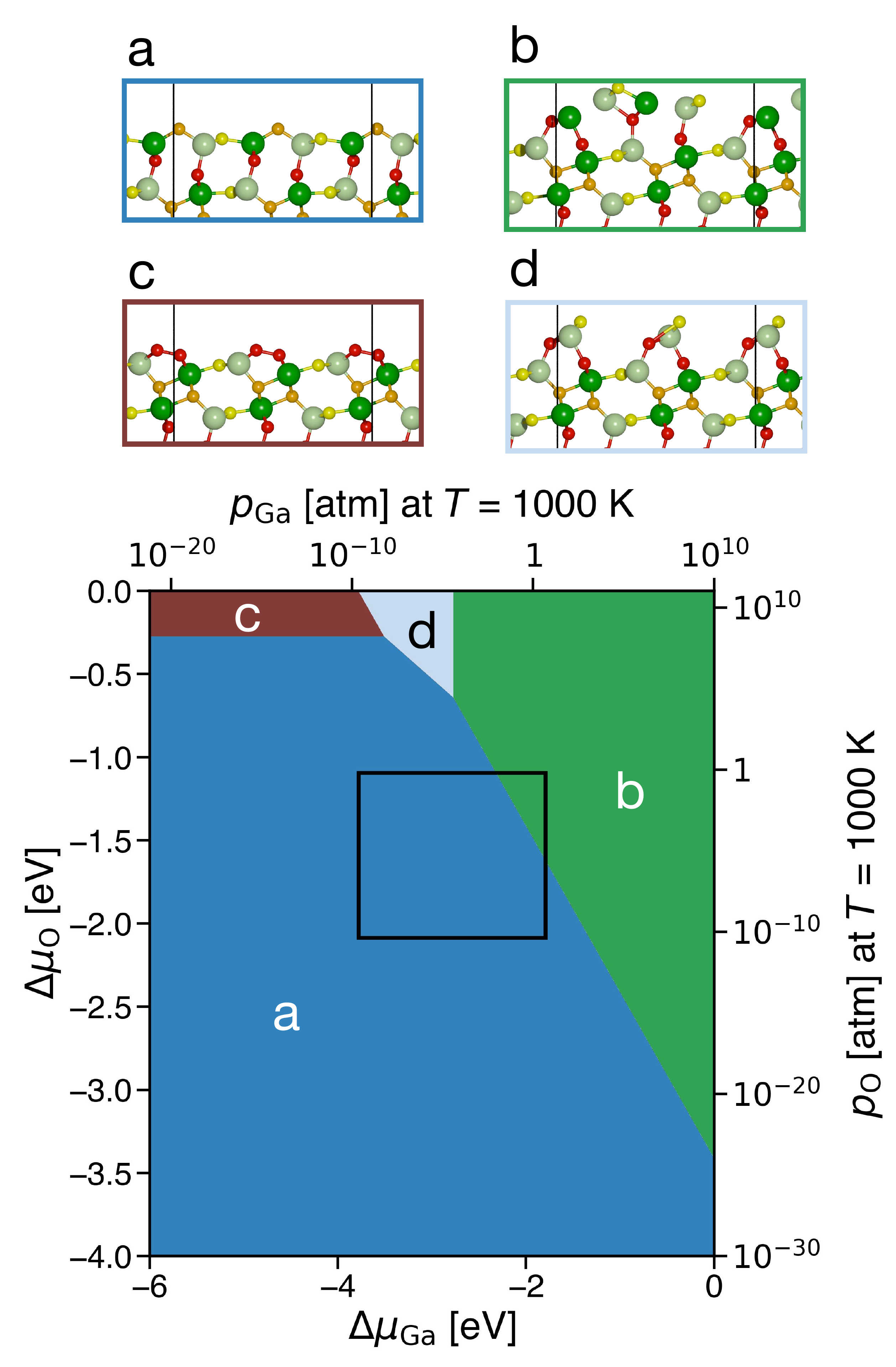}
\caption{Surface phase diagram of $\beta$-Ga$_2$O$_3$(001) obtained with the functional PBEsol, including only the bulk-truncated terminations from Fig.~\ref{fig:bulk_truncated_structures}. The top four panels show the relaxed surface structures along [010], which have a region of stability in the phase diagram. At the top and right axes, the dependence of $\Delta \mu_{\text{O}}$ and $\Delta \mu_{\text{Ga}}$ is transformed into a pressure scale at a fixed temperature of \SI{1000}{\kelvin}. The dashed rectangles indicate the experimentally accessible pressure range of $10^{-10}$ to \SI{1}{\atm}.  \label{fig:bulk_truncated_phase_diagram}}
\end{figure}

We start with the most straight-forward approach and include all bulk-truncated (001) structures. They can be created by shifting the termination plane of the (001) surface between the A and B terminations (see Fig.~\ref{fig:bulk_structure}) and enumerating all resulting inequivalent structures. As a result, we obtain 10 structures, \ie 2 stoichiometric terminations, 4 Ga-rich terminations, and 4 O-rich terminations, which are displayed in Fig.~\ref{fig:bulk_truncated_structures}. Including these structures leads to the phase diagram shown in Fig.~\ref{fig:bulk_truncated_phase_diagram}, where we can see the formation of 3 (meta)stable reconstructions in addition to (001)-B. By comparing these metastable reconstructions, we can already observe an interesting result: In none of the reconstructions Ga or O atoms are absorbed on (001)-B, but only on the (001)-A termination (or equivalently by creating vacancies on the (001)-B termination). We don't explore this aspect in more detail, since it only presents an intermediate result.

\begin{figure}
\centering
\includegraphics[width=0.4\textwidth]{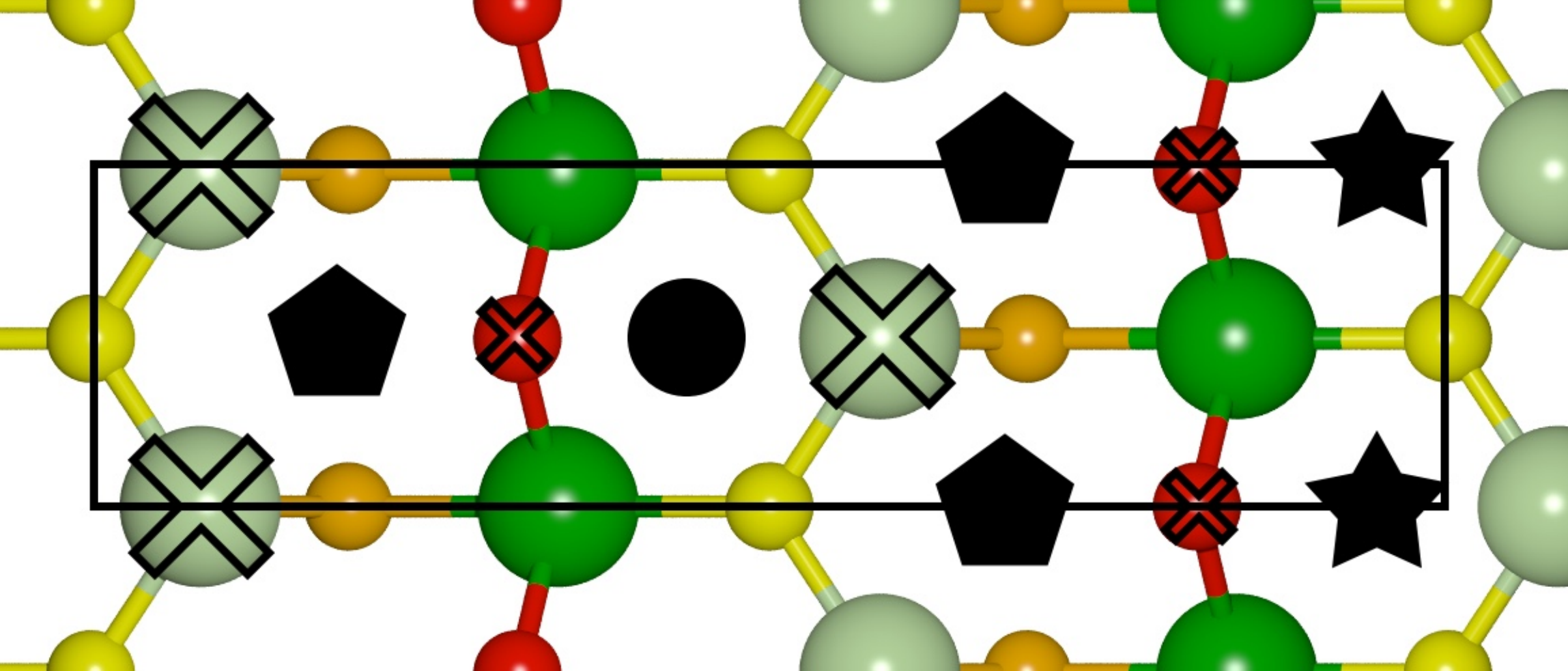}
\caption{Top view of the ideal (001)-A surface. The conventional unit cell is indicated by the black rectangle. Adsorption sites are marked with different symbols, \ie atop sites with unfilled crosses and hollow sites with filled symbols.  \label{fig:adsorption_sites_001}}
\end{figure} 

Another approach is to create adsorption structures based on available adsorption sites in the surface unit cell, as was previously done to study the co-adsorption of Al (or In), Ga, and O on the (010) surface~\cite{wang2021a,wang2020}. Here, we focus on the less stable (001)-A termination that can reconstruct more easily than the B termination. The available adsorption sites (2 atop and 3 hollow sites) are indicated in Fig.~\ref{fig:adsorption_sites_001} in the Appendix. We allow for a maximum of 4 atoms to be adsorbed at the surface of a $1\times 1$ supercell. We filter for the most stable configurations at any given Ga and O composition, and these structures are subsequently relaxed using more accurate computational settings, c.f. \textcolor{black}{Appendix}~\ref{sec:comp_details}. 

The last and most advanced and computationally expensive approach is the replica-exchange grand-canonical (REGC) method~\cite{zhou2019,zhou2022} in conjunction with {\it ab initio} molecular-dynamics (MD) simulations to simulate the (001) surface in contact with a reactive O$_2$ atmosphere. This approach has been shown to yield accurate surface phase diagrams for Si(100) in a hydrogen gas phase~\cite{zhou2022}. The resulting surface phase diagrams include all vibrational contributions, including anharmonic effects, and thus can model realistic $T,p_O$ growth conditions. In REGC-MD, $S$ = $L \times M$ replica of the initial surface structures are considered, each evolving in a different thermodynamical state~($T_l$,$\mu_m$), where $l = 1, 2,\dots, L$ and $m = 1, 2,\dots , M$. During the simulation, the algorithm alternatively (i) diffuses all replicas via \textit{ab initio} MD, \ie $S$ parallel \textit{ab initio} MD runs in an $N V T_l$ ensemble are performed for a fixed number of MD steps or (ii) performs either a particle insertion/removal on all replica in the grand-canonical ensemble or a replica-exchange swap of the $T_l , \mu_m$ variables among pairs of replicas~\cite{zhou2019}. We sample a temperature range from \SIrange{600}{900}{\degreeCelsius} in steps of \SI{100}{\degreeCelsius}, which represents commonly reported temperature values for MOCVD/MOVPE and MBE growth of $\beta$-Ga$_2$O$_3$~\cite{galazka2018,higashiwaki2020}. Chemical potentials are chosen as -0.3, -0.2, -0.14, -0.08, and \SI{-0.02}{\electronvolt}. We, thus, focus on the O-rich regime, where more O atoms can adsorb at the surface to enhance the sampling of possible reconstructions. In total, we simulate 20 different replica for each initial surface structure. The simulation time per replica is \SI{5.0}{ps}~(500 REGC steps).

Ideally, we would simulate the (001) surface in Ga \textit{and} O$_2$ gas phases. While an extension of REGC-MD to a constrained equilibrium, where the gallium and oxygen reservoirs are independent from each other, is straightforward in principle, it would require additional development of an adaptive $\mu_i$ grid to avoid a dimensional explosion~\cite{zhou2019}, which goes beyond the scope of this work. Instead, we run three separate simulations, where the input structures are the (001)-A, the (001)-B, and a Ga-rich structure (structure 9 in Fig.~\ref{fig:bulk_truncated_structures}) in a $1\times 2$ supercell. These different input structures, thus, not only allow for a variation of the initial surface composition, but also correspond to naturally occurring surface structures during growth. After each REGC-MD simulation, we construct a phase diagram based on the number of chemisorbed and physisorbed oxygen atoms. To efficiently estimate the partition function from our REGC sampling, we adopt the multistate Bennett acceptance ratio (MBAR)~\cite{shirts2008} approach, as implemented in the Python package pymbar~\cite{zotero-679}. We note that the acquired phase diagrams only serve as an initial screening of metastable structures. In addition, it is well known that GGA functionals overestimate the binding energy of O$_2$ in its triplet ground state by more than \SI{1}{\electronvolt}~\cite{perdew1992,perdew1996,ernzerhof1997}. Any conclusions drawn from the phase diagrams should therefore be treated with caution. Instead of relying on the REGC phase diagrams, we extract for each ($T_l$,$\mu_m$) state the most stable structures based on their free energy and the number of chemisorbed O atoms. To include them in our aiAT phase diagram, we deposit the reconstructed top layers on the respective ideal bulk-truncated structures and relax the resulting slabs using the same DFT settings as the previous two screening approaches (see also \textcolor{black}{Appendix}~\ref{sec:comp_details}).

\subsection{Experiment}

Scanning transmission electron microscopy~(STEM) images were obtained for a homoepitaxial (001) oriented $\beta$-Ga$_2$O$_3$ sample deposited with In-mediated metal-exchange catalysis (MEXCAT). The images were recorded with a high-angle annular dark-field (HAADF) detector with an inner acceptance angle of \SI{35}{mrad} and a camera length of \SI{196}{mm}. The sample was prepared and studied in the cross-section view perpendicular to the [010] direction. It was deposited at a growth temperature $T_g = \SI{800}{\degreeCelsius}$ with an oxygen flux of \SI{0.75}{sccm}. The beam equivalent pressures of Ga and In were $BEP_{\text{Ga}} = \SI{1.27e-7}{mbar}$ and $BEP_{\text{In}} = \SI{5.2e-8}{mbar}$, corresponding to particle fluxes of $\Phi_{\text{Ga}} = \SI{2.2}{nm^{-2}s^{-1}}$ and $\Phi_{\text{In}} = 1/3 \Phi_{\text{Ga}}$, respectively. Additional details on this particular sample can be found in Ref.~\cite{mazzolini2020}. The MEXCAT allows, under proper growth conditions, to limit the incorporation of large amounts of In inside the deposited layer. \textcolor{black}{On the other hand, we can't exclude an accumulation of In atoms on the surface of the investigated homoepitaxial layer.}

\textcolor{black}{Additionally, a Mg-doped $\beta$-Ga$_2$O$_3$ (001)-oriented bulk crystal obtained from the Czochralski method~\cite{galazka2017} was investigated by RHEED in the same MBE deposition chamber. The substrate was previously chemically etched with H$_3$PO$_4$ and subsequently annealed in a tubular oven in presence of molecular oxygen at \SI{950}{\degreeCelsius} as described in Ref.~\cite{mazzolini2019a}. Diffraction patterns along the [110] and [010] azimuthal directions were collected for the as-prepared sample as well as after an O-plasma treatment at a substrate temperature of \SI{950}{\degreeCelsius} for 15--30 minutes.} 

\section{Results}
\subsection{The (001)-A and (001)-B terminations}
\label{sec:stoichiometric_surfaces}


The (001) surface has two bulk-terminated stoichiometric terminations that are commonly referred to as (001)-A and (001)-B~\cite{bermudez2006}, see also Fig.~\ref{fig:bulk_structure}. The surface structure of the A termination consists of undercoordinated tetrahedral \textcolor{black}{Ga2} and O2 atoms, while the B termination consists of undercoordinated octahedral \textcolor{black}{Ga1} and O1 atoms. Our DFT-PBEsol and DFT-PBE0(0.26) calculations agree with previous results~\cite{bermudez2006,schewski2018,mu2020} that the B termination is significantly more stable than the A termination, as evident from the lower surface free energy $\gamma_{B}=\SI{1.3}{\joule\per\meter^2}<\gamma_{A}=\SI{1.9}{\joule\per\meter^2}$. The surface energies here are calculated for the usual definition of stoichiometric surfaces~\cite{reuter2003}
\begin{equation}
    \gamma_{A/B} = \frac{1}{2A} \left( E_{A,B} - \frac{N_{\text{A/B}}}{N_{\text{bulk}}} e_{\text{bulk}} \,  \right) ,
\end{equation}
where $E_{A,B}$ is the energy of the slab, $N_{A,B}$ is the number of atoms in the slab, and $N_{\text{bulk}}$ is the number of atoms per bulk formula unit. This equation is derived from Eq.~\ref{eq:basic_gamma} under the assumption of thermodynamical stability of the bulk, c.f. Eq.~\ref{eq:bulk_stability}. While both surfaces have the same dangling-bond density and fulfill the electron counting rule~\cite{pashley1989b}, the (001)-A surface cuts Ga2-O2 bonds indicated in light green (cf. the nomenclature for the atoms in Fig.~\ref{fig:bulk_structure}), which are significantly "harder to cut", \ie have larger bond-stretching force constants~\cite{mu2020}, than corresponding Ga1-O1 bonds (same color) that are cut in (001)-B~\cite{mu2020}, which in turn leads to the higher surface energy for the A termination. This also indicates that the A termination is more likely to strongly adsorb atoms. As visualized in Fig.~\ref{fig:001_A_B_truncated_structures}, relaxing both terminations leads to a flattening of the surface, where both Ga1 and Ga2 atoms align horizontally. The atomic displacements decrease strongly beyond the first few atomic layers, and both terminations do not reconstruct upon relaxation. These findings agree with previous results for the (001) surface~\cite{schewski2018,mu2020} and follow the same trend as other surface directions such as (100) and $(\bar{2}01)$~\cite{bermudez2006,schewski2018,mu2020,hinuma2019}.

\subsection{(001) surface phase diagram}
\label{sec:phase_diagram}

\begin{figure*} \centering
\includegraphics[width=1.0\textwidth]{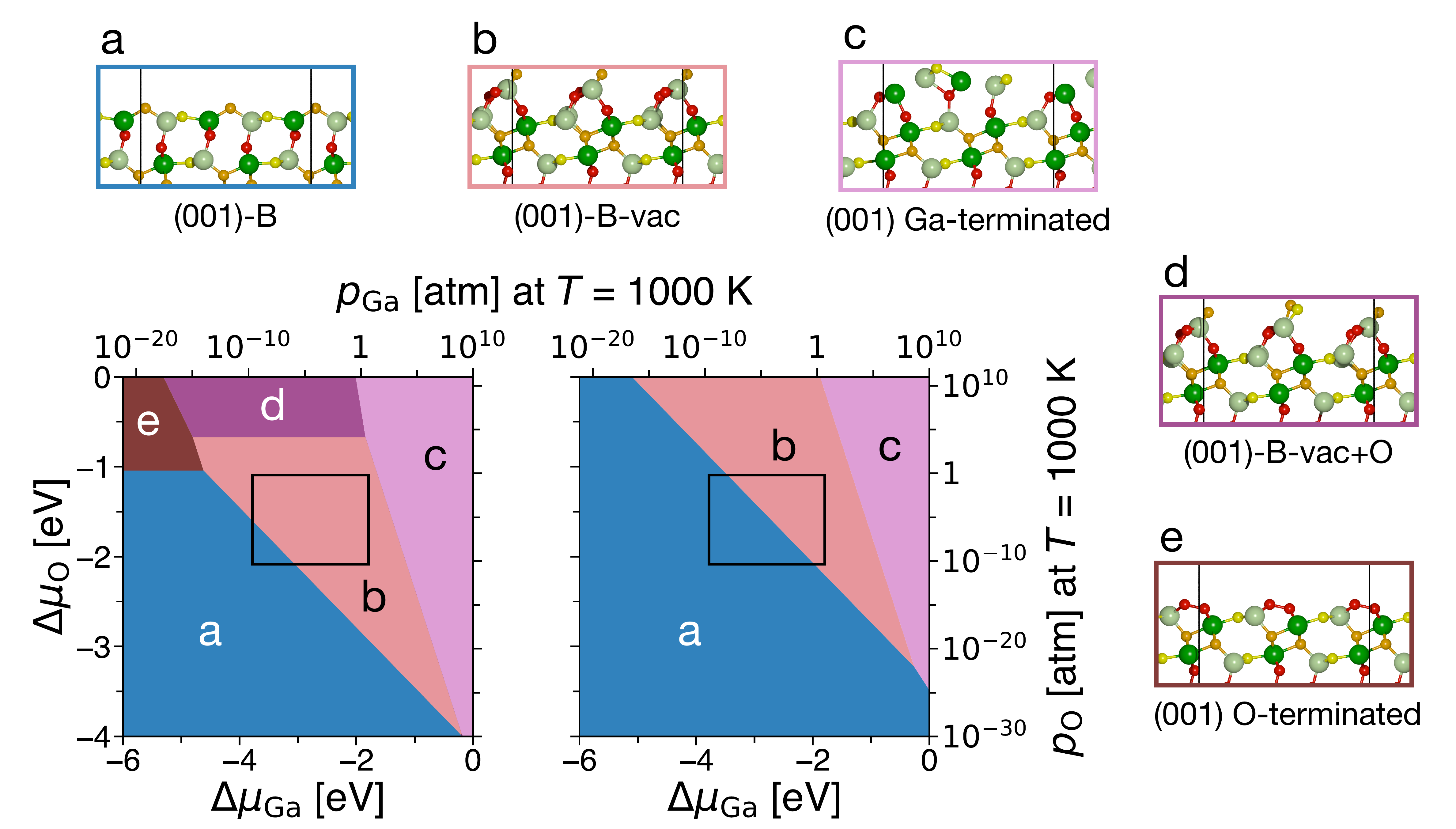}
\caption{Surface phase diagrams of $\beta$-Ga$_2$O$_3$(001) in oxygen and gallium atmosphere obtained with different levels of theory: geometries relaxed with PBEsol (left) and with PBE0(0.26) at the PBEsol geometries (right). Shown are projections along [010] of the surface structures, which have a region of stability in the phase diagram. The chemical potentials $\Delta \mu_{\text{O}}$ and $\Delta \mu_{\text{Ga}}$ are transformed into a pressure scale at a fixed temperature of \SI{1000}{\kelvin} shown as top and right axes. The dashed rectangles indicate the experimentally accessible pressure range of $10^{-10}$ to \SI{1}{\atm}.
\label{fig:001_phase_diagram_strucs}}
\end{figure*} 

\begin{figure} \centering
\includegraphics[width=0.5\textwidth]{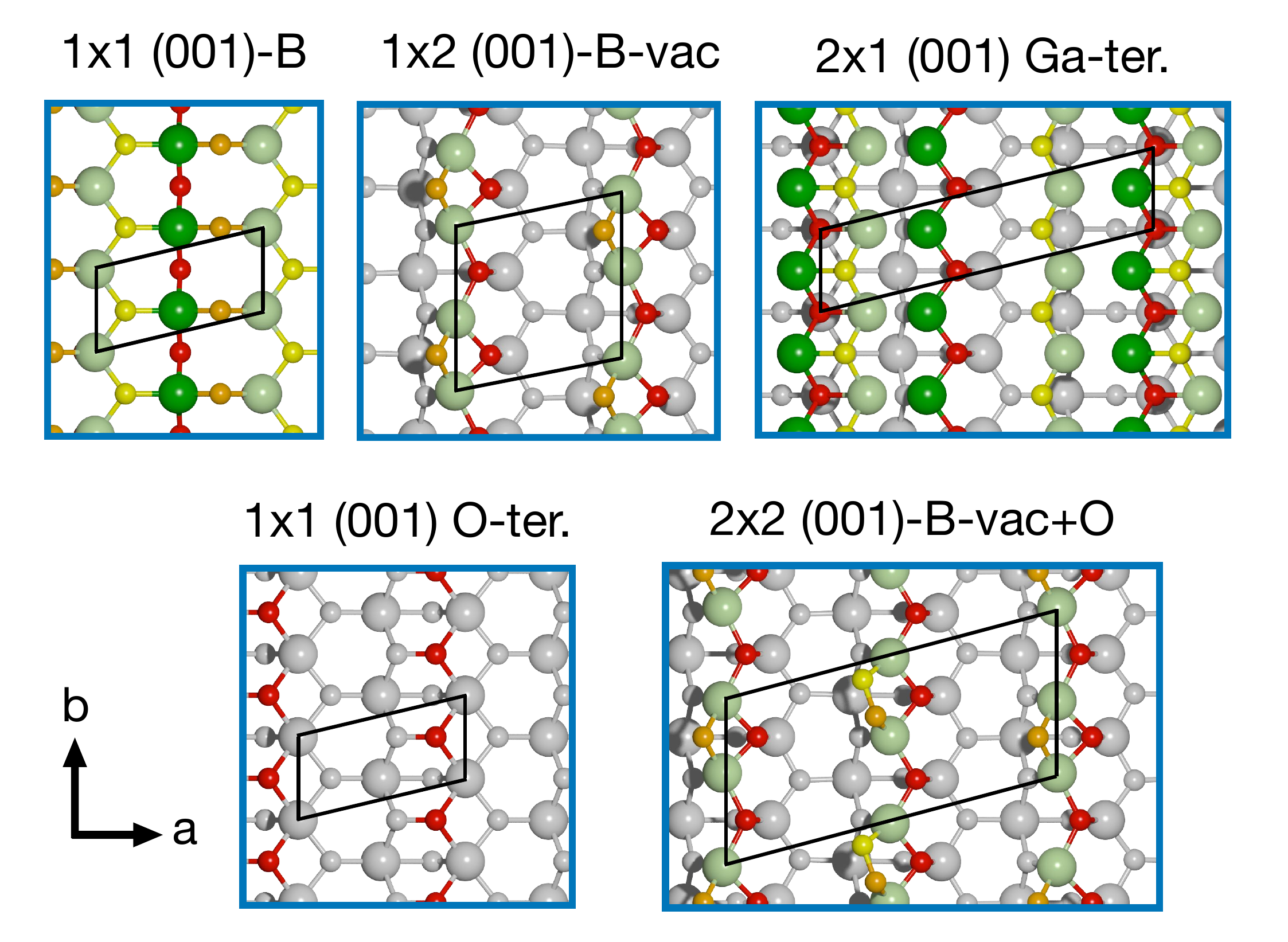}
\caption{\textcolor{black}{Top views of all metastable and stable surface terminations in Fig.~\ref{fig:001_phase_diagram_strucs}. The upper left panel shows the bare (001)-B surface. In the other panels, the adatoms incorporated in the reconstructions are in color, while the atoms of the bare surface are shown in gray. The primitive unit cells are indicated by the black rectangles, with their periodicity indicated in the labeling of the structures. The structures (001) Ga-terminated and (001) O-terminated are referred to as (001) Ga-ter. and (001) O-ter., respectively.}
\label{fig:001_reconstruction_periodicity}}
\end{figure} 

\begin{figure} \centering
\captionsetup[subfigure]{font=small,labelfont=small}
\includegraphics[width=0.5\textwidth]{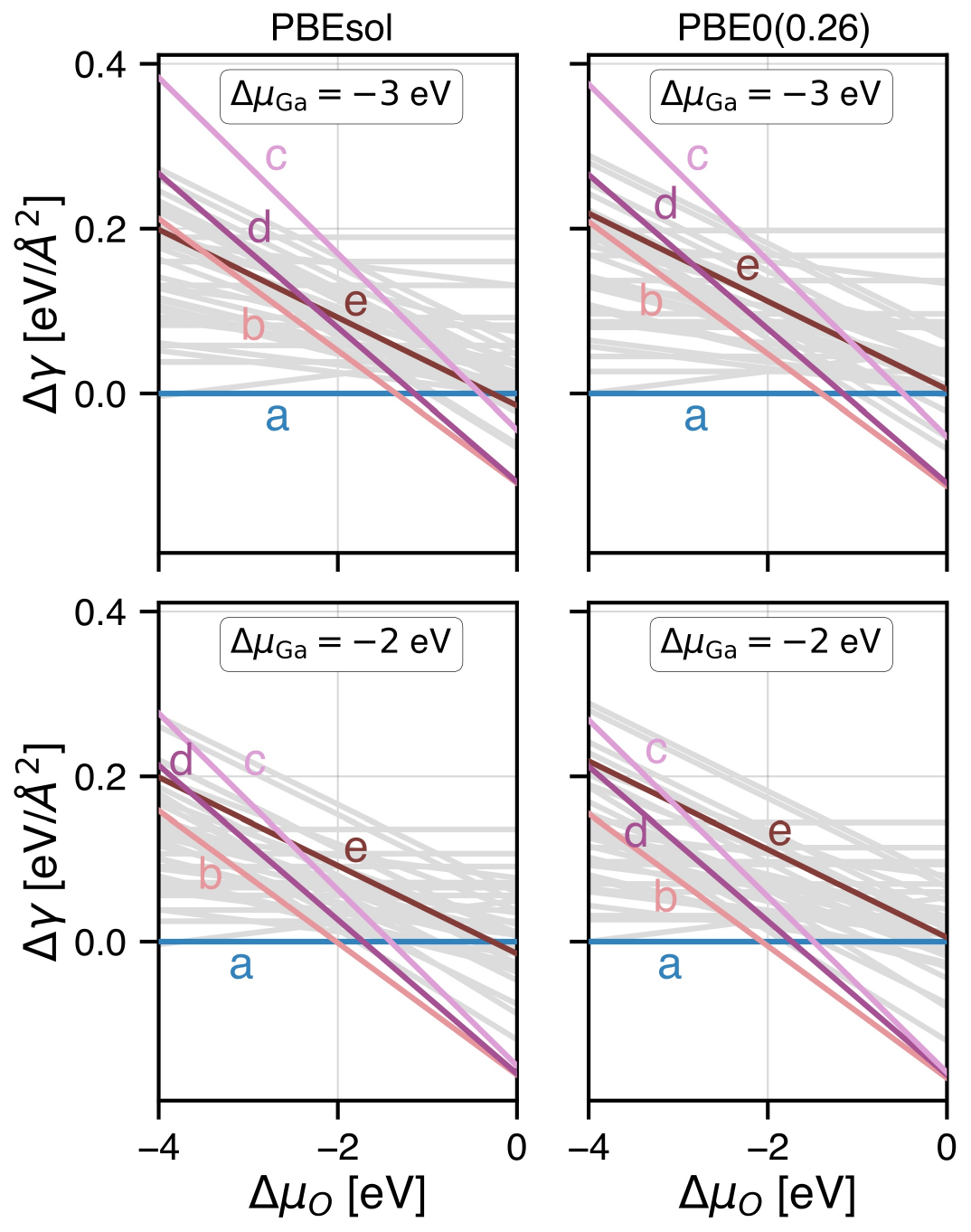}
\caption{Surface free energy difference $\Delta \gamma$ of (001) structures in oxygen and gallium atmosphere obtained with different levels of theory: (left) PBEsol relaxed, (right) PBE0(0.26) at the PBEsol geometry for Ga-poor conditions with $\Delta \mu_{\text{Ga}}=\SI{-3}{\electronvolt}$ (top) and Ga-rich conditions with $\Delta \mu_{\text{Ga}}=\SI{-2}{\electronvolt}$ (bottom). The surface structures with a region of stability in Fig.~\ref{fig:001_phase_diagram_strucs} are highlighted with the same color code, while other surface structures are shown in light gray.
\label{fig:001_phase_diagram_Ga_values}}
\end{figure} 

\begin{table}
\caption{\label{tab:periodictiy}Surface periodicity and composition (with respect to the stoichiometric (001)-A and (001)-B terminations in a $1\times 2$ supercell) of all stable and metastable (001) surface structures. \textcolor{black}{For the definitions of each structure, see the nomenclature in the main text.}}
\begin{tabular*}{\columnwidth}{l@{\extracolsep{\fill}}ll}
  \hline \hline \vspace{-8pt}\\
 & \mc{Composition} & \mc{Periodicity} \vspace{2pt}\\
\hline \\[-8pt]
(001)-B & (001)-B & $1\times 1$ ($0.5 a$ $\times$ $b$)  \vspace{2pt}\\
(001)-B-vac & (001)-A+4Ga+6O & $1\times 2$ ($0.5 a$ $\times$ $2 b$) \vspace{2pt}\\
(001)\,Ga-terminated & (001)-A+8Ga+8O & $2\times 1$ ($a$ $\times$ $b$)\vspace{2pt}\\
(001)-B-vac+O  &  (001)-A+4Ga+7O & $2\times 2$ ($a$ $\times$ $2 b$)\vspace{2pt} \\
(001)\,O-terminated &  (001)-A+4O &  $1\times 1$ ($0.5 a$ $\times$ $b$)\vspace{2pt}\\
\end{tabular*}
\end{table}

\begin{figure}[!h] \centering
\includegraphics[width=0.5\textwidth]{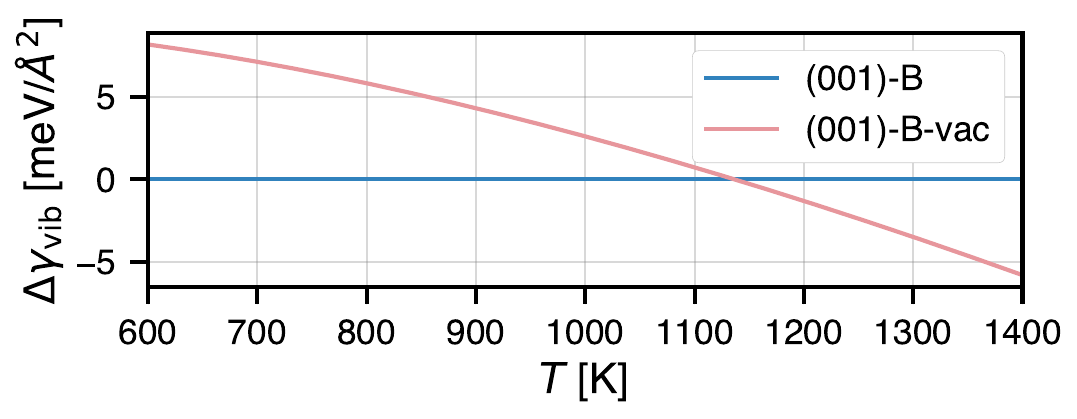}
\caption{Surface free energy contributions from the vibrational free energy $\Delta\gamma_{\text{vib}}(T)$ of the (001)-B and (001)-B-vac structures. Energies are relative to that of the (001)-B surface.
\label{fig:001_vibs}}
\end{figure} 

Let us know focus on possible reconstructions of the (001) surface under realistic conditions. We combine our preselected surface structures and compare their stability in two scenarios: (1) The (001) surface is in contact with O$_2$ and Ga gas phases, which we model as a constrained equilibrium. This decouples the chemical potentials of the two elements and corresponds to the situation found in highly controlled MBE growth. The surface stability is then determined via Eq.~\ref{eq:delta_gamma}. (2) The (001) surface is only in contact with O$_2$ where the Ga chemical potential can be deduced from the assumption that the bulk is thermodynamically stable, cf. Eq.~\ref{eq:bulk_stability}. Here, the surface free energies are only dependent on $\mu_O$ and given by Eq.~\ref{eq:delta_gamma_2}. This would mimic a surface annealing treatment with O$_2$. This second approach is shown in Appendix~\ref{sec:annealing}.

Figure~\ref{fig:001_phase_diagram_strucs} shows the phase diagram of $\beta$-Ga$_2$O$_3$(001) within an atmosphere of oxygen and gallium. The phase diagram on the left side, as obtained by PBEsol, consists of five distinct regions, labeled a to e. An enumeration of the surface periodicity and composition of all found metastable structures is given in Table~\ref{tab:periodictiy}, while a top view of the structures is shown in Fig.~\ref{fig:001_reconstruction_periodicity}. Region a, occupying the lower left half of the diagram, \ie starting from O-poor and Ga-poor conditions, represents the known (001)-B termination.

Between $\Delta\mu_{\text{Ga}}$ values of \SIrange{-4.5}{-0.3}{\electronvolt} and $\Delta\mu_{\text{O}}$ values of \SIrange{-4.0}{-0.6}{\electronvolt}, we identify region b, referred to as (001)-B-vac. As seen in Fig.~\ref{fig:001_reconstruction_periodicity}, this is a $1\times 2$ reconstruction, where two Ga-O tetrahedra are formed as edge-sharing pairs along $b$ by two Ga and three O atoms on top of the (001)-A termination. The Ga atoms thus share one oxygen bond and are separated by a distance of \SI{2.64}{\angstrom}~($\approx 0.87\, b$). Repeated images are perfectly aligned as columns along [010]. Equivalent columns of these tetrahedra are placed along [100] at a distance of 0.5\,$a$. The resulting surface structure is stoichiometric and fulfils the electron counting rule. The smallest supercell in which this reconstruction can form is the $1\times 2$ supercell. This reconstruction can be understood in terms of Ga1 and O3 vacancies~\textcolor{black}{(abbreviated as vac in the following nomenclature)} on the (001)-B termination, hence our label (001)-B-vac.

Under Ga-rich conditions with $\Delta\mu_{\text{Ga}}$ values exceeding \SI{-2.0}{\electronvolt}, we find in region c a $1\times 2$  reconstruction, labeled as (001) Ga-terminated. It is obtained by relaxing the bulk-terminated structure 7 in Fig.~\ref{fig:bulk_truncated_structures}, which is formed by removing the topmost O2 atoms from the (001)-A surface. There is also a separate, less favorable relaxation pathway where solely the topmost Ga2 atoms relax outward, but that reconstruction is about \SI{0.5}{\joule\per\meter^2} higher in surface free energy.

In very O-rich conditions, with $\Delta\mu_{\text{Ga}}$ values exceeding \SI{-1.0}{\electronvolt}, we additionally identify two new regions, d and e. Structure d is a $2\times 2$ reconstruction that is formed by adsorbing an additional O atom at two separate Ga$_2$O$_3$ tetrahedra on the (001)-B-vac structure. As such, we denote it as (001)-B-vac+O. Region e represents a $1\times 1$ oxygen-terminated surface, the (001) O-terminated structure, which corresponds to the bulk-terminated structure 5 in Fig.~\ref{fig:bulk_truncated_structures}. Here, the two exposed O2 atoms form a mutual bond upon relaxation.

Note that while we show the whole phase diagram, MBE growth of Ga$_2$O$_3$(001) is typically performed in a much narrower window of chemical potentials, most commonly at Ga-rich conditions, and high growth temperatures up to \SI{1100}{\kelvin}~\cite{mazzolini2019a}. As such, our phase diagrams also illustrate a more attainable stability region with $T\approx\SI{1000}{\kelvin}$ and $p\approx$ $10^{-10}$ to \SI{1}{\atm}. Here, we only find that the (001)-B structure stabilizes at lower pressures, at approximately $10^{-7}$ to $10^{-10}$ \si{\atm}, and the $1\times 2$ (001)-B-vac reconstruction ar higher pressures. This suggests that (001)-B-vac can indeed stabilize under typical experimental conditions, especially more O-rich and Ga-rich conditions. 

The right side of Fig.~\ref{fig:001_phase_diagram_strucs} shows the phase diagram obtained with PBE0(0.26) using the geometries relaxed with PBEsol. We discern noticeable differences compared to the PBEsol result: Phases a, b, and c are the only constituents of the phase diagram, while phases d and e vanish in O-rich conditions. Within the experimentally accessible conditions, we continue to find only the (001)-B, and the (001)-B-vac reconstruction, but (001)-B exhibits enhanced stability at slightly higher pressures than before.

In addition to the 2D representation of the phase diagram, we also plot the surface free energy differences $\Delta \gamma$ for two fixed values of $\Delta \mu_{\text{Ga}}$ simulating realistic Ga-rich and Ga-poor conditions in Fig.~\ref{fig:001_phase_diagram_Ga_values}. In Ga-poor conditions, we find that two reconstructions, (001)-B-vac (label b) and (001)-B-vac+O (label d), compete closely. Using PBE0(0.26), the surface free energy of all reconstructions changes by an average of only \SI{7}{\milli\electronvolt\per\angstrom^2}, with a maximum increase of \SI{22}{\milli\electronvolt\per\angstrom^2} and a maximum decrease of \SI{8}{\milli\electronvolt\per\angstrom^2}. The relative stability between different reconstructions remains remarkably similar in the PBEsol results. These small changes, however, can have a large impact on the phase diagram when the reconstructions are in close competition near the phase boundaries. This can be seen for example under Ga-poor conditions in the upper right part of Fig.~\ref{fig:001_phase_diagram_Ga_values}, where (001)-B-vac (label b) is stabilized compared to (001)-B-vac+O (label d), leading to (001)-B-vac+O vanishing from the phase diagram in Fig.~\ref{fig:001_phase_diagram_strucs}. Similarly, only (001)-B-vac is stabilizing in O-rich and Ga-rich conditions, as seen in the lower right diagram. 

We now assess the impact of the vibrational free energy on the stability of the newly found reconstructions. Despite being typically small~\cite{reuter2003}, this contribution might play a decisive role in determining the stability of these phases, as shown previously for the example of the polar ZnO(0001) surface in contact with a humid oxygen environment~\cite{valtiner2009}. To this end, we calculate the contribution of the vibrational free energy in the harmonic approximation according to Eq.~\ref{eq:delta_gamma_vib} (details in \textcolor{black}{Appendix}~\ref{sec:comp_details}) using the functional PBEsol. It has been shown previously that LDA, although overbinding, gives good results for the phonon modes in $\beta$-Ga$_2$O$_3$ compared with experiments~\cite{janzen2021}. Since we are mainly interested in their changes compared to the (001)-B reference, calculations at the PBEsol level should be adequate to assess the \textit{differences} in vibrational free energies. Calculations at the PBE0 level to obtain the phonon properties were computationally not affordable.    

Figure~\ref{fig:001_vibs} reveals that the vibrational contributions remain below 15 mJ/m² up to a temperature of 1500 Kelvin. In particular, we find that the bare (001)-B surface gains stability over the (001)-B-vac reconstruction beyond temperatures of 1100 K. These contributions thus appear to be small, and we believe that they don't have a strong influence on the phase diagram except near the phase boundaries. 

\subsection{Reconstructed (001) surface in experiment and theory}
\label{sec:reconstruction}

\begin{figure}[!h] \centering
\includegraphics[width=0.5\textwidth]{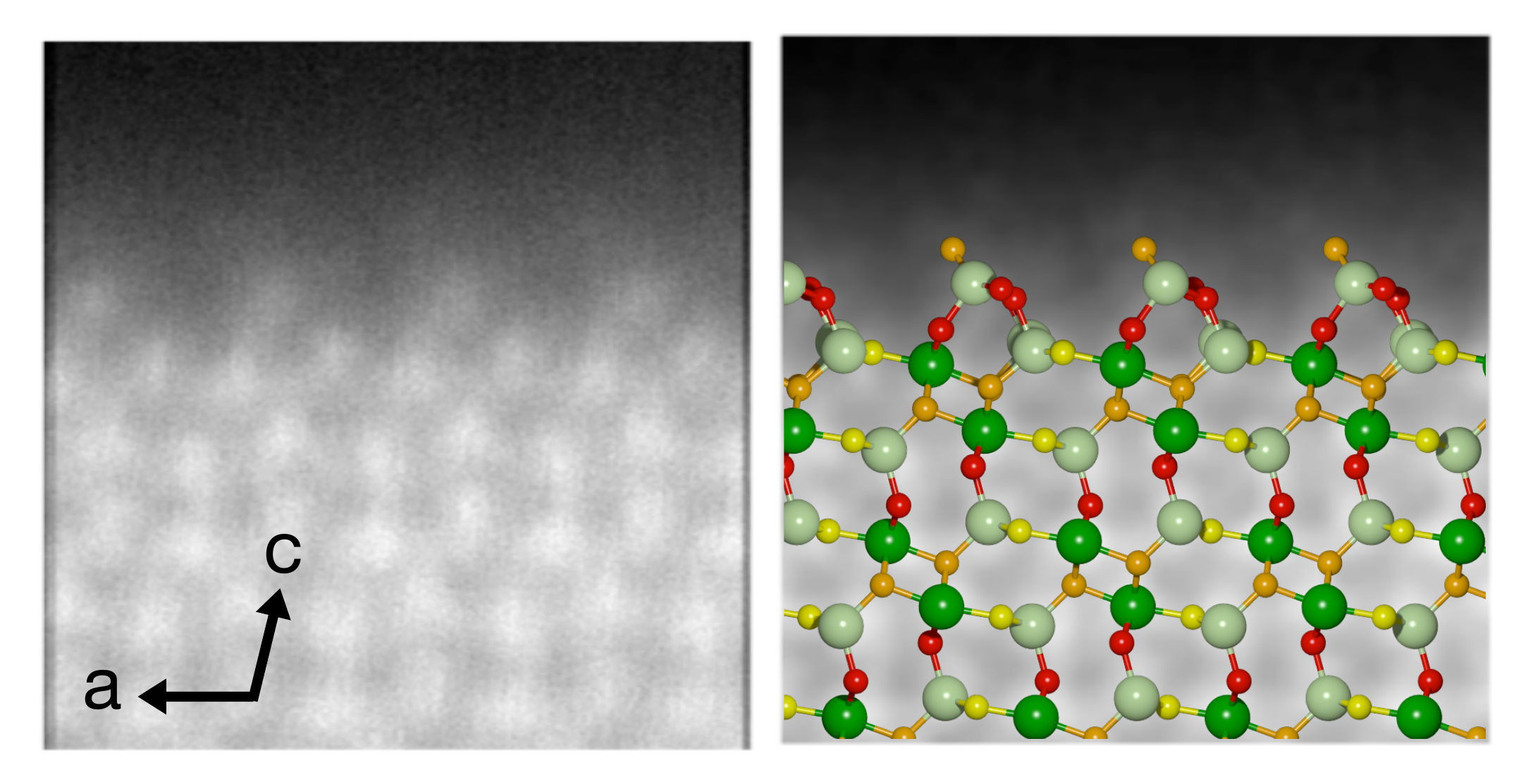}
\caption{HAADF-STEM image of the homoepitaxial (001) layer deposited at $T_g = \SI{800}{\degreeCelsius}$ with an oxygen flux of \SI{0.75}{sccm} with In-mediated MEXCAT. Shown is the projection along [010]. The theoretically predicted (001)-B-vac reconstruction is shown as an overlay on the right.
\label{fig:001_exp_overlay}}
\end{figure} 

\begin{figure} \centering
\captionsetup[subfigure]{font=small,labelfont=small}
\includegraphics[width=0.5\textwidth]{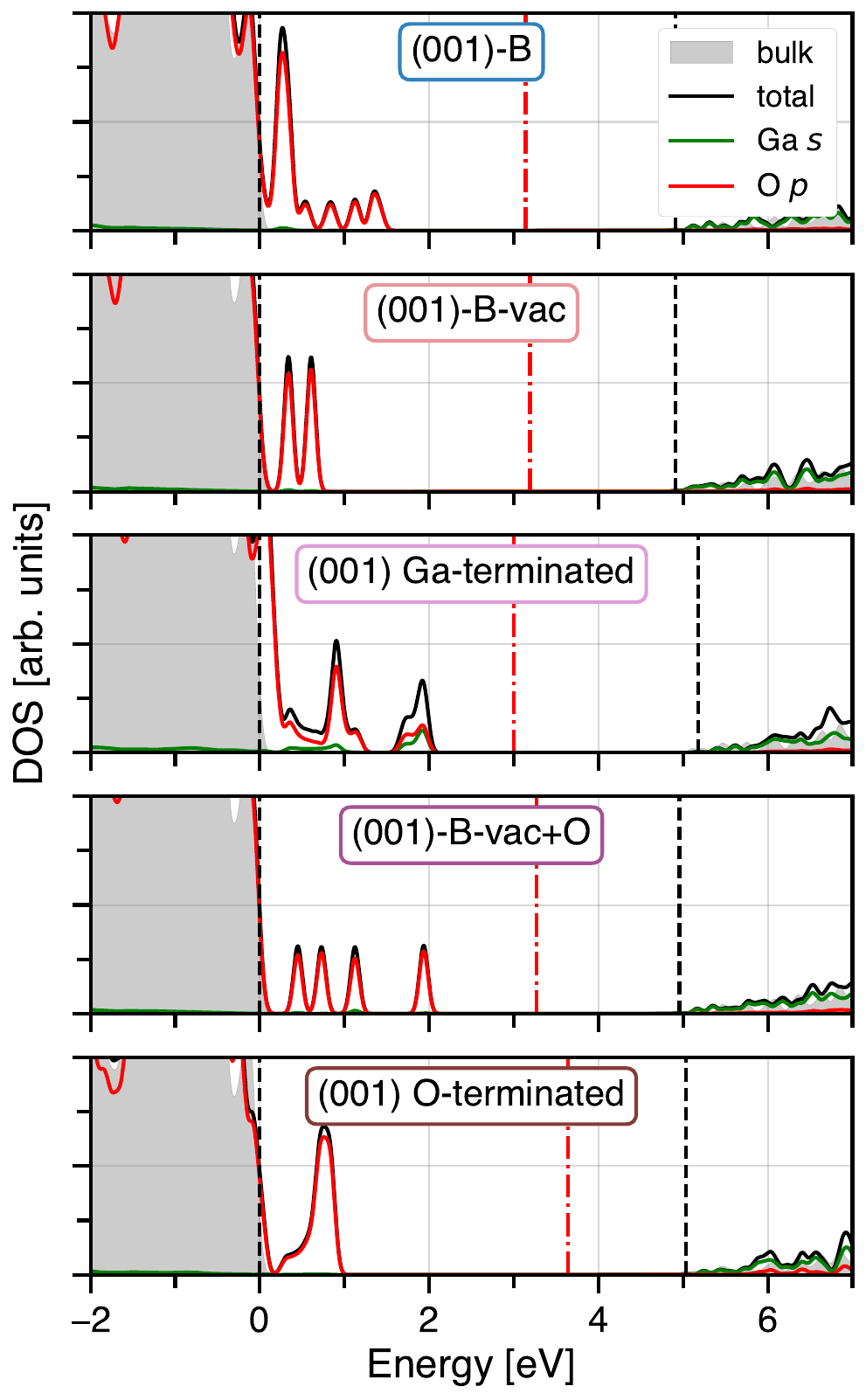}
\caption{DOS of all stable and metastable surface terminations shown in Fig.~\ref{fig:001_phase_diagram_strucs}, obtained with PBE0(0.26) at the PBEsol geometries. The energy is referenced to the VBM. The dashed black vertical lines indicate the VBM and CBM. The dash dotted vertical line indicates the Fermi level. The projections onto the Ga~$s$ states are shown in green, the ones on the O~$p$ states in red. The DOS of the bulk is indicated by the shaded area. \label{fig:DOS_comparison_all_ters}}
\end{figure} 

\begin{table}
\caption{\label{tab:work_gaps}Calculated work functions, $W$, and band gaps, $E_{\text{g}}$, in \si{\electronvolt} for all stable and metastable terminations shown in Fig.~\ref{fig:001_phase_diagram_strucs} obtained with PBEsol and PBE0(0.26) at the PBEsol geometries. The work functions are defined with respect to the VBM.}
\begin{tabular*}{\columnwidth}{l@{\extracolsep{\fill}}llll}
  \hline \hline \vspace{-8pt}\\
   & \multicolumn{2}{c}{PBEsol}
    & \multicolumn{2}{c}{PBE0(0.26)} \vspace{2pt}\\
 & \mc{$W$} & \mc{$E_{\text{g}}$} & \mc{$W$} & \mc{$E_{\text{g}}$} \vspace{2pt}\\
\hline \\[-8pt]
(001)-B & 8.01 & 2.25 & 9.98 & 4.91 \vspace{2pt}\\
(001)-B-vac & 7.54 & 2.20 & 9.46 & 4.91\vspace{2pt}\\
(001) Ga-terminated & 7.01 & 1.96 & 9.01 & 5.18\vspace{2pt}\\
(001)-B-vac+O  &  8.11 & 2.29 & 10.12 & 4.96  \vspace{2pt} \\
(001) O-terminated &  7.78 &  2.38 & 9.76 & 5.03\vspace{2pt}\\
\end{tabular*}
\end{table}

Our HAADF-STEM investigations of homoepitaxially grown (001) layers reveal a clear surface reconstruction that shows excellent agreement with our theoretically predicted (001)-B-vac structure, as shown in Fig.~\ref{fig:001_exp_overlay}. The experimental image shows the formation of distinct Ga adatoms at the surface, with characteristic spacings and orientations matching our simulations.

\begin{figure*} \centering
\captionsetup[subfigure]{font=small,labelfont=small}
\includegraphics[width=1\textwidth]{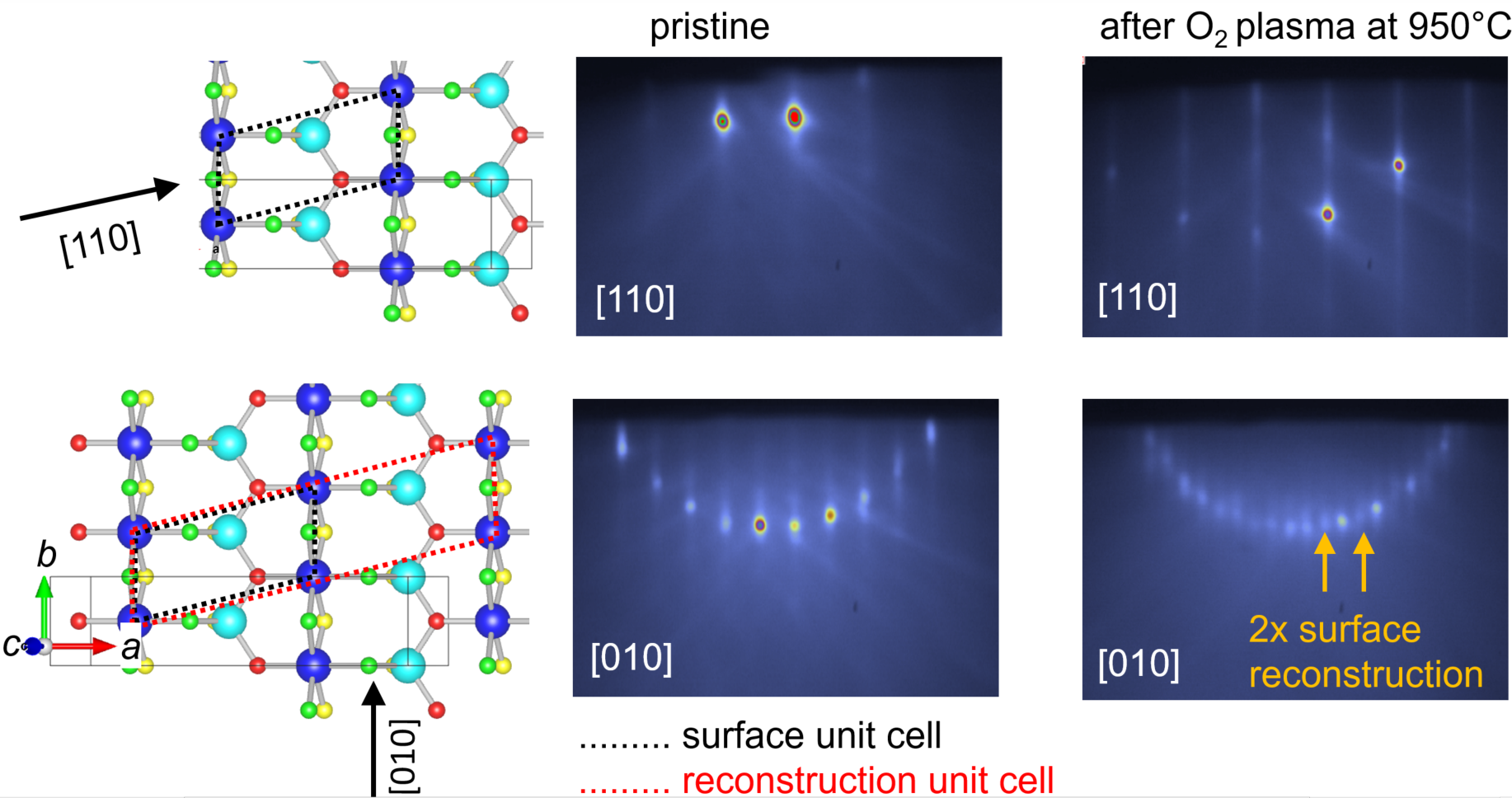}
\caption{RHEED observation of a 2$\times$1 reconstruction along the [010] azimuthal direction. The structures on the left show the (001) surface unit cell, suggested reconstruction unit cell, and related azimuths used for the RHEED analysis. RHEED patterns along the [110] and [010] azimuths are shown for the pristine and reconstructed surface.\label{fig:RHEED2x1}}
\end{figure*}

Interestingly, \textcolor{black}{when focusing on a bare substrate, upon an O-plasma treatment performed at a substrate temperature of \SI{950}{\degreeCelsius}}, our RHEED observations shown in Fig.~\ref{fig:RHEED2x1} indicate a 2$\times$1 reconstruction along the [010] azimuthal direction. This pattern, which persists after cooling and plasma termination, suggests the formation of a different surface reconstruction than the one observed in \textcolor{black}{STEM on the homoepitaxially grown layer with In-mediated MEXCAT}. \textcolor{black}{The reconstruction persists even when the substrate temperature is decreased and the plasma is switched off. The same reconstruction (not shown here) was prepared by heating the same type of substrate that was just solvent-cleaned but not chemically etched and annealed in O-plasma (\SI{1}{sccm} O$_2$ flow, \SI{300}{\watt} RF plasma power, corresponding pressure in growth chamber $\sim \SI{3e-6}{mbar}$) at \SI{950}{\degreeCelsius} for approx. 15 min.} While this indicates the possibility of additional stable surface configurations under O-rich conditions, particularly at elevated temperatures, the exact nature of this reconstruction requires further investigation. Our calculations predict a 2$\times$1 Ga-terminated reconstruction under extremely Ga-rich conditions, but there is no current evidence that this is the same structure as observed in the RHEED images. The experimental conditions during O-plasma treatment suggest that a different reconstruction mechanism may be involved.

Having determined which reconstructions can be stabilized, we explore their properties and the impact on potential applications. The electronic structure in the form of the density of states (DOS) of the five metastable structures is shown \textcolor{black}{in Fig.~\ref{fig:DOS_comparison_all_ters}}. The work functions and band gaps are listed in Table~\ref{tab:work_gaps}. All of these terminations exhibit pronounced surface states near the valence band. Except for the (001) Ga-terminated one, all reconstructions have an excess oxygen on the surface. Their dangling bonds are filled, and the excess electrons are localized in the band gap. \textcolor{black}{For these oxygen-rich reconstructions, the filled gap states are predominantly of O~$2p$ character ($>$90\%), arising from oxygen dangling bonds. In contrast, the (001) Ga-terminated surface exhibits a qualitatively different electronic structure, with the gap states showing approximately 35\% Ga~$4s$/$4p$ character compared to only $\sim$7\% for the oxygen-rich reconstructions, reflecting the distinct bonding environment of undercoordinated surface Ga atoms.} The band gaps of all surface structures are slightly larger than those of the bulk system, most likely due to quantum confinement. The values of (001)-B and (001)-B-vac are nearly identical, while the work function of (001)-B-vac is \SI{0.5}{\electronvolt} smaller than that of (001)-B. Oxygen adsorption and formation of the (001)-B-vac+O reconstruction increases the work function above that of the bare surface by \SI{0.1}{\electronvolt}. The systematic trends in work functions, decreasing from O-rich to Ga-rich terminations, follow the expected behavior based on the ionic model~\cite{noguera2000,tasker1979}.

The pronounced surface states near the valence band may impact carrier transport and band alignment in devices. The presence of filled oxygen dangling bonds in most reconstructions, except the Ga-terminated surface, suggests that these surfaces may act as hole traps. This could significantly impact the carrier dynamics in $p$-type devices, similar to effects observed for other orientations~\cite{mu2020}. We also note that the moderate reduction in work function ($\approx$ \SI{0.5}{\electronvolt}) and the nearly identical band gaps of (001)-B-vac and (001)-B suggests that the reconstruction maintains the favorable band alignment while providing improved structural stability.

\subsection{Possible reconstructions with In}

\begin{figure} \centering
\captionsetup[subfigure]{font=small,labelfont=small}
\includegraphics[width=0.5\textwidth]{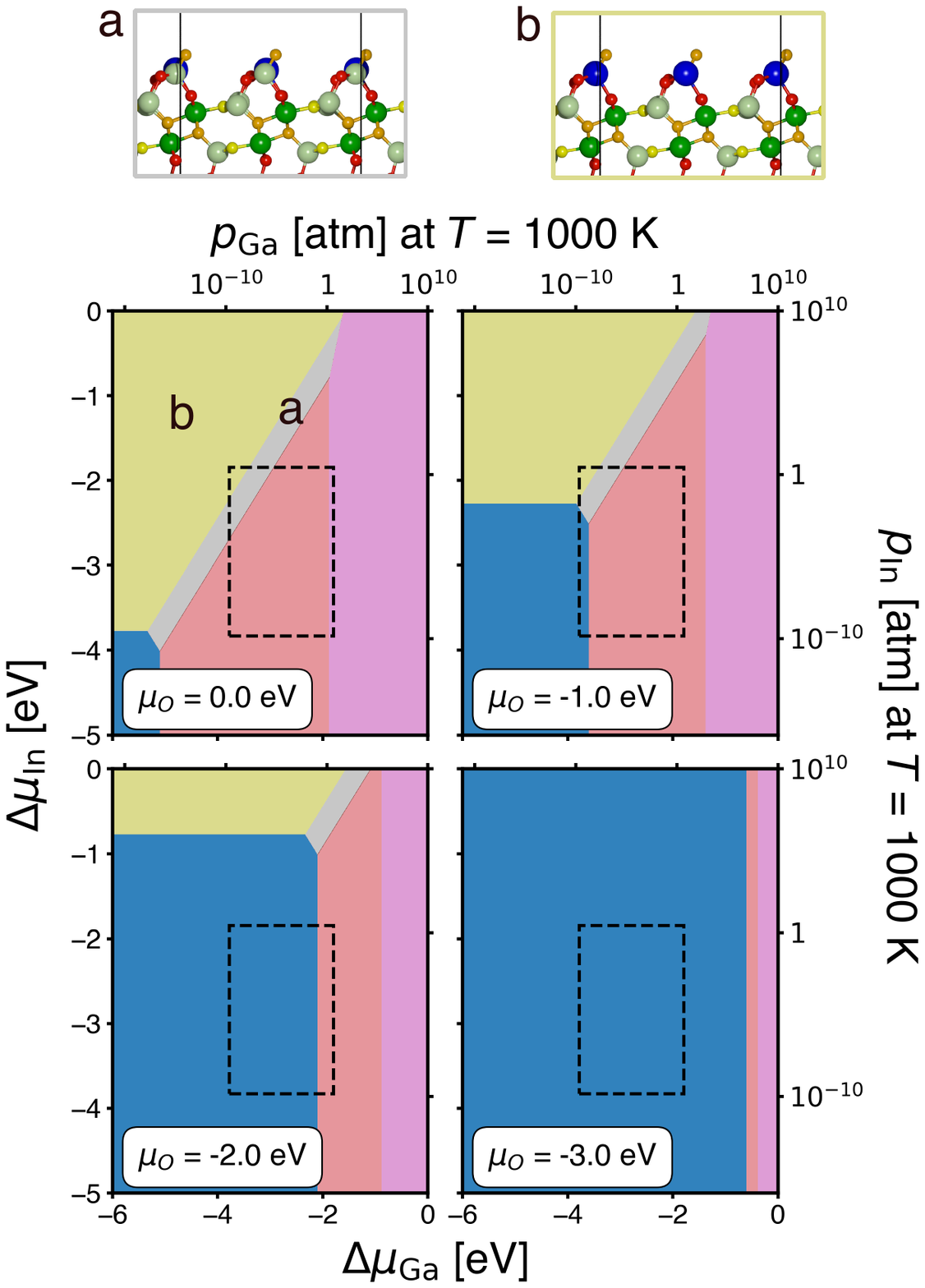}
\caption{Surface phase diagrams of $\beta$-Ga$_2$O$_3$(001) in oxygen, indium, and gallium atmosphere obtained with PBE0(0.26) at the PBEsol geometries for different fixed values of $\Delta \mu_{\text{O}}$. Shown are projections along [010] of the surface structures including In atoms (in blue), which have a region of stability in the phase diagram. The color code for the other stable surface structures is the same as in Fig.~\ref{fig:001_phase_diagram_strucs}. The chemical potentials $\Delta \mu_{\text{In}}$ and $\Delta \mu_{\text{Ga}}$ are transformed into a pressure scale at a fixed temperature of \SI{1000}{\kelvin}. The dashed rectangles indicate the experimentally accessible pressure range of $10^{-10}$ to \SI{1}{\atm}.\label{fig:001_phase_diagram_strucs_In}}
\end{figure} 

Having identified the (001)-B-vac reconstruction as a stable surface structure that agrees well with our experimental STEM-HAADF observations (Fig.~\ref{fig:001_exp_overlay}), we also consider the role of In during MEXCAT growth. Given that In is intentionally introduced during the growth process as a metal-exchange catalyst, an important question arises \textcolor{black}{whether part of the In atoms segregate at the surface or fully desorb during the catalytic process}. Therefore, we investigated the possibility that some or all of the surface Ga atoms in the (001)-B-vac reconstruction might actually be substituted by In atoms during growth.

To explore this scenario, we substituted the topmost surface Ga atoms in the (001)-B-vac reconstruction with In atoms at different ratios (0\%, 25\%, 50\%, 75\%, and 100\%). The resulting phase diagrams (Fig.~\ref{fig:001_phase_diagram_strucs_In}) reveal that structures with 50\% (region a) and 100\% (region b) In substitution have small but distinct regions of stability in the experimentally accessible region marked by the dashed rectangle, while intermediate compositions (25\% and 75\%) show negligible stability windows. This binary preference for either half or full In occupation suggests a cooperative effect in In substitution, where the presence of some In atoms facilitates the incorporation of additional In atoms to the surface layer up to these specific ratios.

The surface phase diagrams also reveal a strong dependence of In incorporation at the surface layer on the oxygen chemical potential. Under O-rich conditions ($\Delta\mu_\text{O}=\SI{0.0}{\electronvolt}$), In-containing reconstructions are stable in the upper-left portion of the phase diagram, particularly in the region where both $\Delta\mu_\text{In}$ and $\Delta\mu_\text{Ga}$ are moderately negative ($\Delta\mu_\text{In}>\SI{-3.0}{\electronvolt}$ and $\Delta\mu_\text{Ga}<\SI{-2.0}{\electronvolt}$). This suggests that In preferentially occupies surface sites when sufficient oxygen is available to form stable In-O bonds.

As conditions become increasingly O-poor (moving from $\Delta\mu_\text{O}=\SIrange{0.0}{-3.0}{\electronvolt}$), the stability region for In-containing surface structures progressively shrinks until it completely disappears at $\Delta\mu_\text{O}=\SI{-3.0}{\electronvolt}$. At medium to high oxygen partial pressures, In can thus be incorporated at the surface layer and remain stable, while low oxygen partial pressures lead to a destabilization of In-containing surface layers. This behavior aligns with experimental observations during In-mediated MEXCAT growth, where optimal growth conditions typically require moderate oxygen pressures~\cite{williams2024, karg2023, vogt2021}.

\section{Summary and Conclusions} 

Through a comprehensive combination of first-principles calculations and experimental probes, we have systematically investigated surface reconstructions of $\beta$-Ga$_2$O$_3$(001). Our \textit{ab initio} thermodynamics approach, combined with REGC-MD simulations, revealed several stable surface reconstructions depending on the chemical environment. The most notable finding is the stoichiometric (001)-B-vac reconstruction, which shows remarkable stability across a wide range of experimental conditions, and agrees well with HAADF-STEM images of a homoepitaxially grown (001) layer. The (001)-B-vac structure is a 1$\times$2 reconstruction where two Ga-O tetrahedra are formed as edge-sharing pairs along $b$ by two Ga and three O atoms on top of the (001)-A termination, or alternatively could be formed by Ga1 and O3 vacancies on the (001)-B termination. This formation of tetrahedra at the surface is also found in other oxides, such as the stoichiometric $\sqrt{31}$ reconstruction studied recently for Al$_2$O$_3$(0001)~\cite{hutner2024} and for $\alpha$-Fe$_2$O$_3$(0001)~\cite{redondo2023}. The prevalence of such tetrahedral motifs suggests a general tendency of oxides to minimize surface energy through the formation of fully coordinated metal-oxygen polyhedra. We find no stabilized terminations with oxygen excess, again supporting similar previous findings in Al$_2$O$_3$(0001)~\cite{hutner2024}.

The phase diagrams demonstrate that under typical MBE growth conditions ($T \approx \SI{1000}{\kelvin}$, $p \approx \SIrange{e-10}{1}{\atm}$), two main terminations dominate: the well-known (001)-B structure at lower oxygen pressures ($\SI{e-7}{\atm}$ to $\SI{e-10}{\atm}$) and the (001)-B-vac reconstruction at higher pressures. The stability of both terminations is confirmed by both PBEsol and PBE0(0.26) calculations, with only minor quantitative differences between the two methods. The calculated electronic properties show that all structures exhibit pronounced surface states near the valence band.

Our investigation of In-mediated growth revealed an interesting cooperative effect in In substitution, where 50\% and 100\% In-substituted structures show distinct stability regions, while intermediate compositions are energetically unfavorable. This binary preference suggests a complex interplay between surface energetics and atomic arrangements during MEXCAT growth. The stability of In-containing reconstructions shows a strong dependence on the oxygen chemical potential, with maximum stability under O-rich conditions.

This study offers fundamental insights into the surface chemistry of $\beta$-Ga$_2$O$_3$(001) and provides a comprehensive understanding for optimizing growth conditions in MEXCAT processes. Future investigations should address several open questions arising from this work. One is to find the kinetic pathways between different surface structures, particularly the transition mechanism from (001)-B to (001)-B-vac. The observation that tetrahedral surface motifs appear in our found reconstructions and across multiple oxide systems in literature can potentially lead to predictive design rules for engineering oxide surface properties in the future.

\section{Acknowledgements}
This work was carried out in the framework of GraFOx, a Leibniz-Science Campus partially funded by the Leibniz Association. The authors gratefully acknowledge the Gauss Centre for Supercomputing e.V. (\url{www.gauss-centre.eu}) for providing computing time on the GCS Supercomputer SuperMUC-NG at the Leibniz Supercomputing Centre (\url{www.lrz.de}). Part of the computations were performed on the HPC systems Raven and Cobra at the Max Planck Computing and Data Facility. We thank Yuanyuan Zhou for help in setting up and analyzing the REGC calculations. K.L. gratefully acknowledges discussions with Matthias Scheffler on the early phases of this work, the setup and interpretation of the surface calculations, and financial support by his Grant TEC1P by the European Research Council 
(ERC) Horizon 2020 research and innovation programme, Grant No. 740233. \textcolor{black}{The input and output files needed to reproduce the results are openly available at NOMAD~\cite{draxl2018, draxl2019,scheidgen2023} with the following link:~\cite{zotero-item-3326}.}

\appendix
\counterwithin{figure}{section}
\counterwithin{table}{section}

\section{Computational details \label{sec:comp_details}}

\begin{figure*}[htb] \centering
\includegraphics[width=0.9\textwidth]{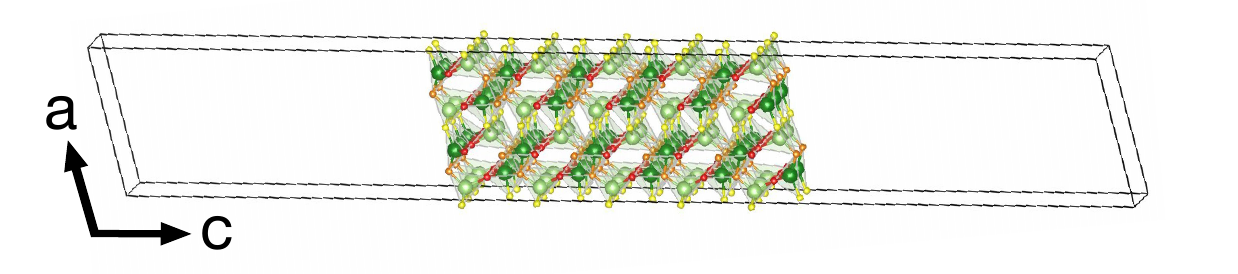}
\caption{\textcolor{black}{Complete $1\times 2$ supercell used for the surface calculations, shown for the (001)-B termination. The slab consists of 5 bulk layers and is embedded in at least \SI{50}{\angstrom} of vacuum to prevent spurious interactions between periodic images. The view is along the [010] direction. The color scheme follows Fig.~\ref{fig:bulk_structure}: octahedrally coordinated Ga1 atoms are shown in dark green, tetrahedrally coordinated Ga2 atoms in light green, and the three inequivalent oxygen species O1, O2, and O3 are shown in orange, red, and yellow, respectively.}
\label{fig:supercell}}
\end{figure*}

All DFT calculations are performed using the all-electron FHI-aims package~\cite{blum2009,abbott2025}. Exchange-correlation effects are treated (i) in the generalized gradient approximation by the revised PBE functional for solids and surfaces (PBEsol)~\cite{perdew2008}, which has shown high accuracy in determining the lattice parameters and elastic properties of Ga$_2$O$_3$~\cite{wouters2020,lion2022}
and (ii) with the hybrid functional PBE0~\cite{adamo1999,ernzerhof1999} including \SI{26}{\percent} exact exchange [denoted as PBE0(0.26)], which matches the experimental gap of \SI{4.9}{\electronvolt}~\cite{tippins1965, orita2000,deak2019}. The calculated bulk lattice parameters obtained by using PBEsol are $a=\SI{12.28}{\angstrom}$, $b=\SI{3.05}{\angstrom}$, $c=\SI{5.81}{\angstrom}$, and $\beta=\SI{103.72}{\degree}$, which agree very well with the experimental values~\cite{ahman1996} of $a=\SI{12.21}{\angstrom}$, $b=\SI{3.04}{\angstrom}$, $c=\SI{5.82}{\angstrom}$, $\beta=\SI{103.82}{\degree}$. The surface structures are created using pymatgen~\cite{ong2013,sun2013} and are modeled as slabs in the supercell approach, including at least \SI{50}{\angstrom} of vacuum. We simulate the (001) surface in $1\times 1$ and $1\times 2$ supercells of the conventional cell, including 5 bulk layers, each consisting of 20 atoms. (Note that all surface phase diagrams in this work are shown for the $1\times 2$ slabs\textcolor{black}{, see Fig.~\ref{fig:supercell}}.) The resulting slabs are symmetric due to the inversion symmetry of the bulk structure, allowing us to extract the properties of the single surface. By defining the surfaces using the conventional bulk cell, the resulting surface unit cell is twice the size of the primitive cell, \ie doubled along the $c$ direction, see also Fig.~\ref{fig:bulk_structure}. The Brioullin zone of the $1\times 1$ slab is sampled with a $4\times 12 \times 1$ k-grid and scaled accordingly for larger supercells. All atoms in the slab are relaxed with the final forces being smaller than \SI{E-3}{\electronvolt\per\angstrom}. In the PBEsol calculations, the basis set and numerical grids are defined by \textit{tight} settings for both the gallium and oxygen atoms, while for the PBE0(0.26) calculations we employ \textit{intermediate} settings. These settings ensure convergence of the surface energies to \SI{0.01}{\joule\per\meter^2}. As the bulk geometries obtained with the two functionals are comparable, it is not necessary to use a hybrid functional for relaxation and to obtain reliable surface energies. The surface free energies on the right side of Fig.~\ref{fig:001_phase_diagram_O2} differ by less than \SI{0.03}{\joule\per\meter^2} upon relaxing all internal coordinates of the slabs using PBE0(0.26).

Vibrational free energies for selected structures are calculated in the harmonic approximation using the Python package Phonopy~\cite{togo2023a,togo2023b}. We simulate the $1\times 2$ slabs in a $1\times 2$ supercell, resulting in lateral dimensions of $\SI{12.2}{\angstrom} \times \SI{12.2}{\angstrom}$, previously used in defect calculations of $\beta$-Ga$_2$O$_3$~\cite{deak2017}. The respective energies are obtained using PBEsol and \textit{tight} basis settings with a k-grid of $6 \times 6 \times 1$.

The REGC simulations are performed using the python package FHI-panda~\cite{zhou2019,fhipanda}. The input structures are the three different relaxed bulk-truncated structures 1, 2, and 9 in Fig.~\ref{fig:bulk_truncated_structures}. Each slab contains 80 atoms (4 bulk-layers), where the bottom half is fixed, and the bottom surface is passivated by pseudo-hydrogen. The surface regions, where atoms can be inserted or removed, are defined as the regions \SI{5}{\angstrom} along the surface normal starting \SI{2}{\angstrom} below the topmost surface atoms. This allows for the possible insertion or removal of sub-surface sites during the simulation. In each REGC step, the probability of performing a replica-exchange step is set to \SI{10}{\percent}. DFT calculations are performed using PBEsol and \textit{light} basis settings with a tail correction for the van der Waals interactions, computed with the Tkatchenko-Scheffler scheme~\cite{tkatchenko2009}. The Brioullin zone is sampled with a $4\times 5 \times 1$ k-grid. The {\it ab initio} MD simulations to diffuse all replicas are performed in a canonical ($NVT$) ensemble for \SI{0.01}{\pico\second} and using a timestep of \SI{1}{\femto\second}, resulting in \SI{5}{\pico\second} long trajectories for each replica. The $NVT$ ensemble is sampled using the stochastic-velocity rescaling thermostat~\cite{bussi2007}. The system is confined to a sphere of \SI{5}{\angstrom} by applying a repulsive polynomial potential of order four via PLUMED~\cite{bonomi2009}.

\section{Surface stability in oxygen gas \label{sec:annealing}}

\begin{figure*} \centering
\includegraphics[width=0.95\textwidth]{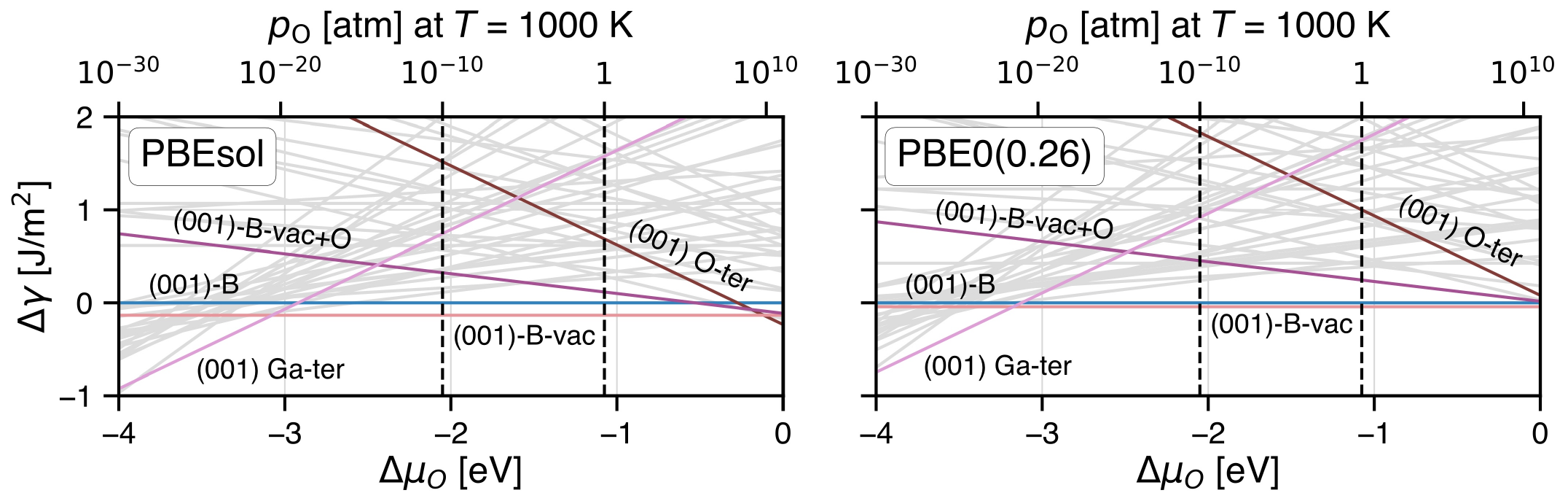}
\caption{Surface free energies for all investigated (001) structures in oxygen atmosphere. The values are obtained using the functional PBEsol (left) and PBE0(0.26) at the PBEsol geometries (right). \textcolor{black}{Energies are referenced to the (001)-B surface}. The chemical potential $\Delta \mu_{\text{O}}$ is transformed into a pressure scale at a fixed temperature of \SI{1000}{\kelvin} (top axes). Structures with a window of stability in Fig.~\ref{fig:001_phase_diagram_strucs} are highlighted by color. Other surface structures are shown in light gray. The dashed vertical lines indicate the experimentally accessible pressure range from $10^{-10}$ to \SI{1}{\atm}.
\label{fig:001_phase_diagram_O2}}
\end{figure*} 

\begin{figure}[!h] \centering
\includegraphics[width=0.5\textwidth]{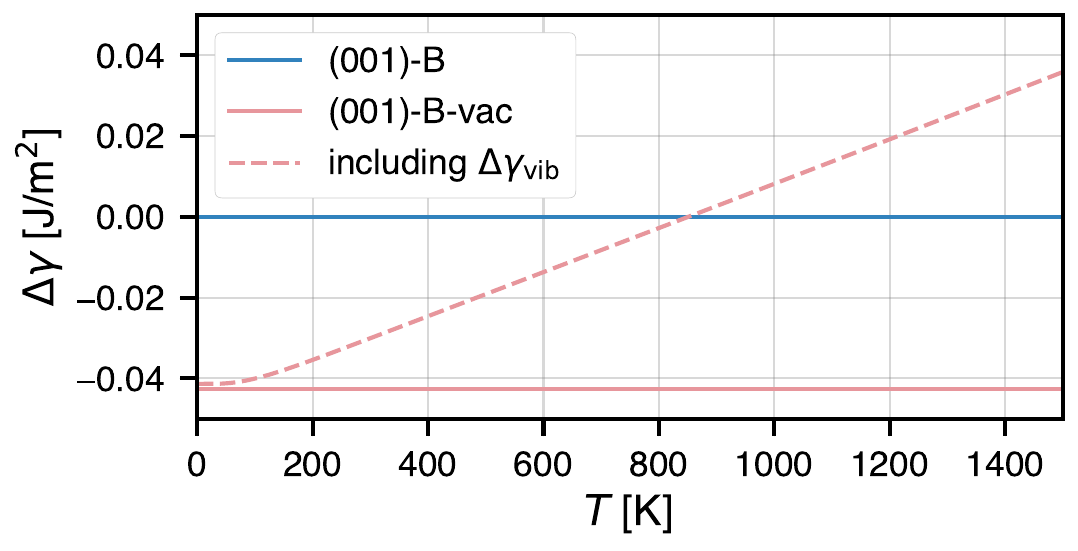}
\caption{Surface free energies of the (001)-B and (001)-B-vac surfaces in oxygen atmosphere using PBE0(0.26). Energies are referenced to the (001)-B surface. The dashed line includes the contribution of the vibrational free energy, $\Delta\gamma$, in the harmonic approximation obtained by PBEsol and determined from Eq.~\ref{eq:delta_gamma_vib_2}.
\label{fig:001_vibs_O2}}
\end{figure} 

We can also consider the surface in contact only with an oxygen reservoir. If we further consider the surface to be in equilibrium with the underlying bulk and the gas phase, then the chemical potentials in Eq.~\ref{eq:delta_gamma} are related by~\cite{qian1988,reuter2003,reuter2005a,rogal2007} 
\begin{align}
    e_{\text{bulk}} = 2\,\mu_{\text{Ga}} + 3\,\mu_{\text{Ga}} \, , \label{eq:bulk_stability}
\end{align}
with $e_{\text{bulk}}$ being the total energy of a bulk Ga$_2$O$_3$ formula unit that approximates the corresponding Gibbs free energy. We can then rewrite Eq.~\ref{eq:delta_gamma} by substituting $\mu_{\text{Ga}}$ using Eq.~\ref{eq:bulk_stability}:
\begin{equation} 
\begin{split}
    \Delta \gamma & (T,p_{\text{O}}) = \frac{1}{2A} \bigg[ E_{\text{slab}}-E_{\text{ref}} \\
    & -\left(\Delta N_{\text{O}}-\frac{3}{2}\,\Delta N_{\text{Ga}}\right)\,\mu_{\text{O}}( T,p_{\text{O}} )-\frac{1}{2}\Delta N_{\text{Ga}}\,e_{\text{bulk}} \bigg] \, . 
\end{split}
\label{eq:delta_gamma_2}
\end{equation}
Finally, when vibrational contributions are considered as part of the Gibbs free energy, we obtain an additional term contributing to the surface free energy:
\begin{equation} 
\begin{split}
    \Delta \gamma_{\text{vib}} & (T) = \frac{1}{2A} \bigg[ F^{\text{vib}}_{\text{slab}}(T)-F^{\text{vib}}_{\text{ref}}(T) -\frac{1}{2}\Delta N_{\text{Ga}}\,f^{\text{vib}}_{\text{bulk}}( T) \bigg] \, .
\end{split}
\label{eq:delta_gamma_vib_2}
\end{equation}

Figure~\ref{fig:001_phase_diagram_O2} showcases the resulting surface free energy using PBEsol and PBE0(0.26), with particular emphasis on the previously identified metastable structures. It is noticeable that the O-rich structures have higher surface energy when obtained with PBE0(0.26). Where previously the (001) O-terminated reconstruction stabilized in O-rich conditions, now only the stroichiometric (001)-B-vac reconstruction remains stable. In the Ga-rich limit, the (001) Ga-terminated is now predicted to be stable below $\Delta \mu_{\text{O}}<\SI{-3.2}{\electronvolt}$. Within the experimentally accessible range (defined by conditions prevalent in typical experimental annealing or MBE growth processes), the terminations (001)-B and (001)-B-vac are separated by less than \SI{5}{\milli\joule\per\meter^2}. 

The close energetic proximity of these structures suggests strong competition and possible metastability in this region. The often (small and) overlooked contributions of vibrational free energy and configurational entropy could play a major role in determining the relative stability. To this end, we calculate the vibrational contributions to the surface free energy in the harmonic approximation for the most stable stoichiometric terminations as shown in Figure~\ref{fig:001_vibs_O2}. The vibrational contributions remain below \SI{10}{\milli\joule\per\meter^2} up to a temperature of \SI{1500}{\kelvin}. Specifically, the bare (001)-B surface gains stability over the (001)-B-vac reconstruction beyond temperatures of \SI{850}{\kelvin}. It is worth noting, however, that these energy differences are so small that other factors, such as the configurational entropy, may further influence this stability. Consequently, a comprehensive understanding of these reconstructions may require further analysis.

To summarize, we predict that varying the annealing conditions can result in different (001) surface morphologies, (1) the $2\times 1$ (001) Ga-terminated reconstruction in extremely O-poor (Ga-rich) conditions, (2) the $1\times 2$ (001)-B-vac reconstruction, for temperatures below \SI{850}{\kelvin}, and (3) the bare (001)-B termination, for temperatures above \SI{850}{\kelvin}, in more O-rich (Ga-poor) conditions. However, the energetic differences are so small that both phases can and likely will occur simultaneously.

\clearpage
%

\end{document}